\documentclass[twocolumn]{aastex62}
\usepackage{epstopdf}
\usepackage{xcolor}
\usepackage{amsmath}
\usepackage{verbatim}
\usepackage{natbib}
\usepackage{romannum}
\newcommand{\angstrom}{\mbox{\normalfont\AA}}

\graphicspath{{./}{figures/}}

\shorttitle{Stellar MZR and Mass-loading factor}
\shortauthors{Leethochawalit et al.}

\begin{document}

\title{Evolution of the Stellar Mass--Metallicity Relation. II. Constraints on Galactic Outflows from the Mg Abundances of Quiescent Galaxies}

\author[0000-0003-4570-3159]{Nicha Leethochawalit}
\affiliation{Cahill Center for Astronomy and Astrophysics,
California Institute of Technology, Pasadena, CA 91125, USA}
\affiliation{National Astronomical Research Institute of Thailand, Mae–Rim, Chiang Mai, 50180, Thailand}

\author[0000-0001-6196-5162]{Evan N. Kirby}
\affiliation{Cahill Center for Astronomy and Astrophysics,
California Institute of Technology, Pasadena, CA 91125, USA}

\author{Richard S. Ellis}
\affiliation{Department of Physics and Astronomy,
University College London, Gower Street, London WC1E 6BT, UK}

\author{Sean M. Moran}
\affiliation{Smithsonian Astrophysical Observatory,
60 Garden Street, Cambridge, MA 02138, USA}

\author{Tommaso Treu}
\affiliation{Department of Physics and Astronomy,
University of California, Los Angeles, CA 90095, USA}
\accepted{September 24, 2019}
\begin{abstract}
We present the stellar mass--[Fe/H] and mass--[Mg/H] relation of quiescent galaxies in two galaxy clusters at $z\sim0.39$ and $z\sim0.54$. We derive the age, [Fe/H], and [Mg/Fe] for each individual galaxy using a full-spectrum fitting technique. By comparing with the relations for $z\sim0$ SDSS galaxies, we confirm our previous finding that the mass--[Fe/H] relation evolves with redshift. The mass--[Fe/H] relation at higher redshift has lower normalization and possibly steeper slope. However, based on our sample, the mass--[Mg/H] relation does not evolve over the observed redshift range. We use a simple analytic chemical evolution model to constrain average outflow that these galaxies experience over their lifetime, via the calculation of mass-loading factor. We find that the average mass-loading factor $\eta$ is a power-law function of galaxy stellar mass, $\eta \propto M_*^{-0.21\pm0.09}$. The measured mass-loading factors are consistent with the results of other observational methods for outflow measurements and with the predictions where outflow is caused by star formation feedback in turbulent disks. 
\end{abstract}


\section{Introduction} \label{sec:intro}

Over the past five decades, we have made significant progress in measuring gas metallicities in star-forming galaxies. The relation between galaxy luminosity (stellar mass) and interstellar oxygen abundances in extragalactic \ion{H}{2} regions was established 50 years ago \citep[e.g.][] {McClure1968, Lequeux1979, Garnett1987}. \citet{Tremonti2004} used the large statistical sample size of local star-forming galaxies from the Sloan Digital Sky survey (SDSS) and confirmed that the correlation between metallicity and mass is more fundamental than that between metallicity and luminosity. The mass--metallcity relation (MZR) is such that more massive galaxies have higher gas metallicities. In later work, the gas-phase MZR was also found to be present at high redshift and also to evolve with redshift \citep[e.g.][]{Maiolino2008, Zahid2013}.

Although the gas-phase MZR has been known for almost four decades, its physical drivers are still debated. Early works mostly suggested that galactic winds are the primary agent that drives the relation \citep[e.g.,][]{MathewsBaker1971, Larson1974, Garnett2002, Tremonti2004}. Lower mass galaxies have shallower potential wells and therefore can retain less of the metals they produced \citep{DekelSilk1986}. However, later works argued that the metallicity is regulated by a more complex mechanism that includes the interplay between inflow, outflow, and enrichment rate \citep[e.g.][]{FinlatorDave2008, Dave2011}. \citet{Spitoni2010} attempted to use chemical evolution models to explain the observed gas-phase MZR in the SDSS galaxies. They found that a range of models that include either outflow only, both inflow and outflow, or inflow with variable outflow can all explain the observed data equally well. In essence, there is a degeneracy between inflow and outflow rate that cannot be differentiated using only the data from the gas-phase MZR.

However, metals do not reside only in gas. Especially at lower redshift $z<0.5$, the cold gas fraction is $<20\%$ in most star-forming galaxies and less than a few percent in quiescent galaxies \citep[e.g.][]{Gobat2018}. The majority of disk and metal mass is in stars \citep[e.g.,][]{Werk2014}. When \citet{Gallazzi2005} measured the stellar metallicities of local SDSS galaxies using spectral indices, they found that stellar metallicity also exhibits a tight correlation with stellar mass in galaxies with $M_*>10^9~M_\odot$. \citet{Kirby2013} measured stellar metallicities of individual stars in the Local Group and found that the correlation extends down to dwarf galaxies with stellar mass as low as $10^3~M_\odot$.     

In principle, measuring stellar metallicity in addition to gas-phase metallicity can break the degeneracy between inflow and outflow \citep[e.g.,][]{Lu2015}, resulting in better constraints on chemical evolution models. However, recent works that have attempted to incorporate both stellar and gas-phase MZRs found that it is difficult to reconcile the two MZRs. \citet{Lian2018} found that the stellar metallicities of local galaxies are generally lower than expected based on their gas-phase metallicities. The discrepancies are larger for lower mass galaxies. The only models that can reconcile both MZRs have to invoke either a steep initial-mass function (IMF) slope (almost twice the slope of the \citealt{Salpeter1955} IMF) or a strong outflow (ejection of all metals produced) at early times.

Determining stellar metallicities in star-forming galaxies is challenging and subject to potentially large biases. First, emission lines from the interstellar medium have to be either subtracted or modeled together with stellar absorption lines. Second, even if the emission lines are modeled perfectly, the measurements of age and metallicity are still subject to bias, especially in the spectra of young stellar populations. The bias generally makes the population appear older and more metal poor than the true values \citep[e.g.,][]{Leethochawalit2018} by as large as $\sim0.5$ dex if the priors are not treated carefully \citep{Ge2018, CidFernandes2018}.

Quiescent galaxies provide an alternative and a more convenient way to constrain galaxy chemical evolution models. The populations are older. The contamination from emission lines is less concerning. The spectral analysis is overall less prone to systematic biases. Additionally, the (near) absence of gas in quiescent galaxies simplifies chemical evolution models by eliminating the need to consider metals in the gas. 

In this paper, we continue the work of \citet[][hearafter Paper I]{Leethochawalit2018} in quantifying the stellar MZR of quiescent galaxies as a function of redshift. We expand the sample size, from galaxies in a galaxy cluster at $z=0.39$ to include an additional sample in a galaxy cluster at $z=0.54$. We measure their magnesium [Mg/H] abundances in addition to iron [Fe/H] abundances. While Fe is produced by both Type \Romannum{2} and Type \Romannum{1}a supernovae, Mg is an alpha element that is mainly produced by Type \Romannum{2} supernovae. Mg therefore has a shorter recycling timescale than Fe. By measuring the abundance of Mg, we now have an indication of the abundance of metals that is approximately instantaneously recycled, which can be used in simple chemical evolution models.

The highlight of this work is that we propose an archaeological method to constrain galactic outflow, in terms of average mass-loading factors that quiescent galaxies experienced as a function of mass, via the measurements of $\alpha$ element abundances. We build upon the technique introduced in Paper I to trace quiescent galaxies back to their epochs of formation using their ages. We conclude that while the [Fe/H] abundance at a fixed galaxy mass appears to evolve with redshift, the [Mg/H] MZR changes neither with the redshift of formation nor with the redshift of observation.

\section{Data} \label{sec:data}
\begin{figure}
\includegraphics[width=0.4725\textwidth]{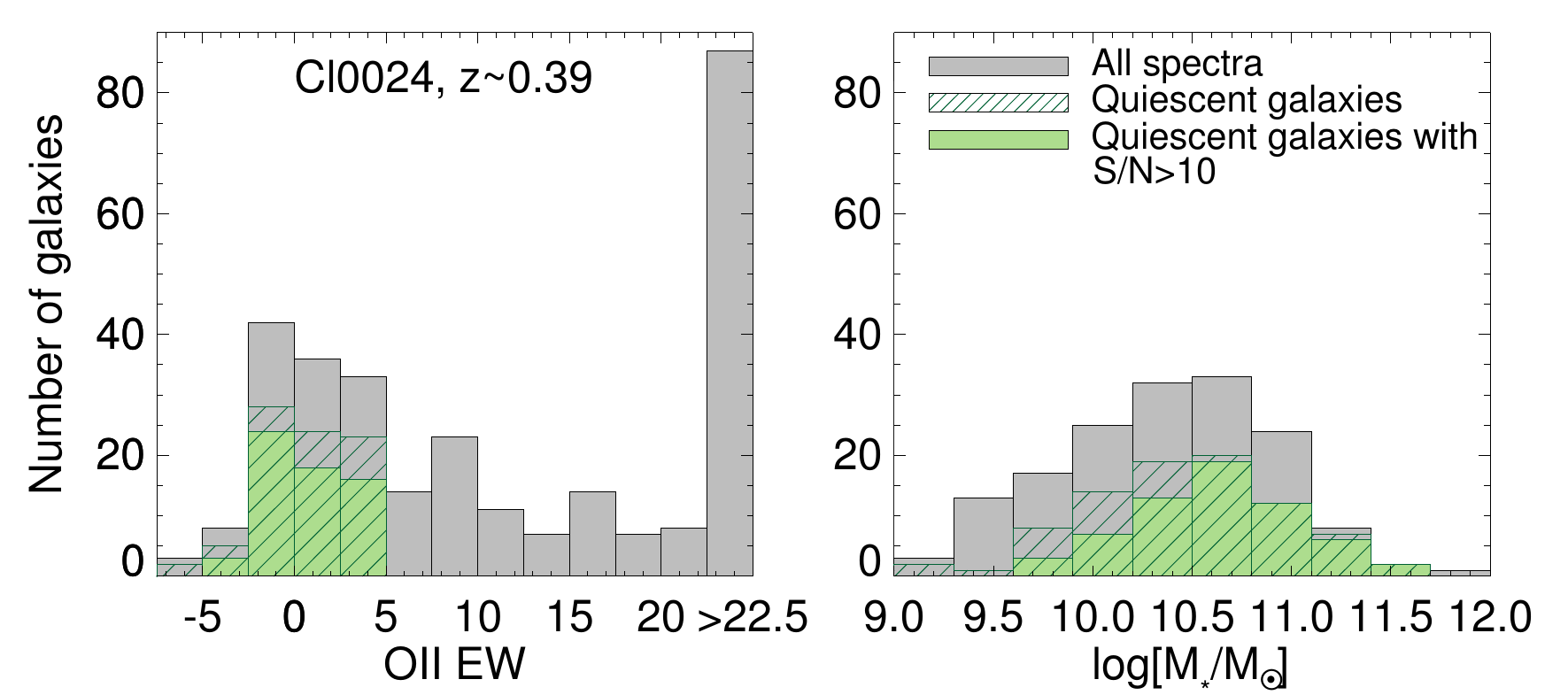}\\
\includegraphics[width=0.4725\textwidth]{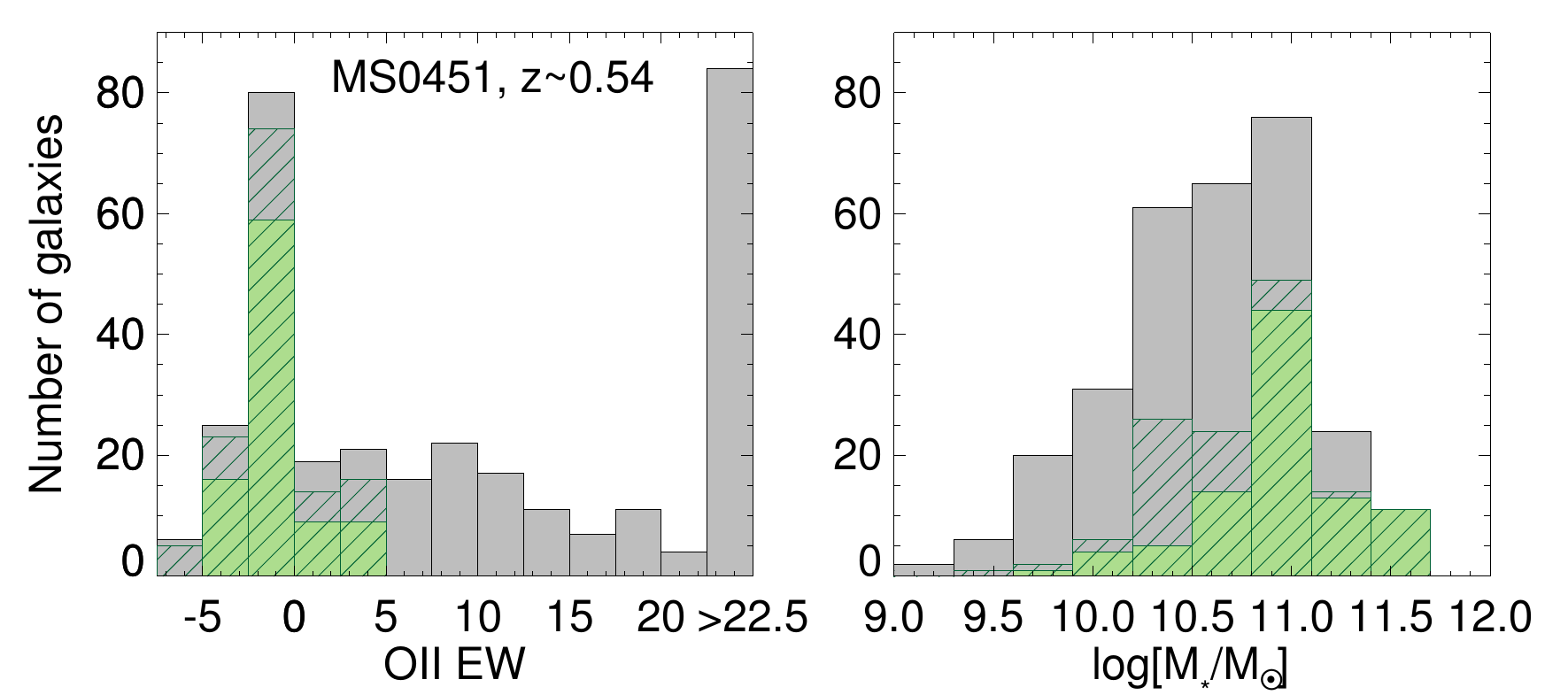}
\caption{Histograms of the parent and selected sample in this paper. We select quiescent galaxies (defined in Sec.\ \ref{sec:data}) with S/N greater than 10 \angstrom$^{-1}$ in the rest frame to be in our sample (solid green). \label{fig:datademo}}
\end{figure}
In Paper I, we presented a stellar MZR based on quiescent galaxies from the galaxy cluster Cl0024+17 at $z=0.39$. In this paper, we expand the sample by including additional quiescent galaxies from the galaxy cluster MS0451 at $z=0.54$. 

The two galaxy clusters comprised the survey of \citet{Moran2007}. The survey provides UV to near-infrared imaging and ground-based optical spectroscopy of member galaxies up to $\sim10$ Mpc in diameter centered on both clusters. The available photometric bands are NUV and FUV from the Galaxy Evolution Explorer (GALEX) satellite, \textit{BVRI} bands from the 3.6 m Canada-France-Hawaii telescope (Cl0024) and the Subaru 8 m telescope (MS0451), F814W ($\sim I$ band) from Hubble Space Telescope, and J and $K_s$ bands from the WIRC camera on the Palomar/Hale 200'' telescope. Using the same method as in Paper I, we use this photometry to estimate the stellar masses of the sample with the SDSS \texttt{KCORRECT} software version v4\_3 \citep{Blanton2007}.

The spectroscopic data for both Cl0024 and MS0451 were obtained with the DEIMOS spectrograph \citep{Faber2003} on the Keck II Telescope. Most data are part of the original survey. We obtained additional DEIMOS spectroscopy to enhance the S/N of a subset of the galaxy spectra in MS0451. In both clusters, the spectra were obtained with 1'' wide slits with a spectral resolution of at least $R=2000$, spanning rest-frame wavelengths from $\sim3500$ to 6000 \angstrom. The detailed spectroscopic observations of Cl0024 are described by \citet{Moran2005} and Paper I.

We now summarize the previous and new DEIMOS spectroscopy in MS0451. In 2003, \citet{Moran2005} observed 11 slitmasks for 1 hr each with the 600 line mm$^{-1}$ grating centered at 7500 \angstrom\@. Based on these initial data, the study identified cluster members and performed a deeper follow-up in 2004--2005 with 10 additional slitmasks with the same grating centered at 6800 \angstrom\@. The integration time was 2.5--4 hrs per slitmask. In total, the original survey identified 319 member galaxies in MS0451. \citet{Moran2007} gives further details of these observations. In December 2016 and October 2017, we additionally observed one more slitmask with the same grating centered at 7200 \angstrom\ for a final follow-up. In the mask design, we prioritized previously observed quiescent galaxies whose spectra combined over all observing runs could achieve S/N of at least 8 \angstrom$^{-1}$ within 5 additional hours of integration time. The actual integration time was 4.5 hrs.

All DEIMOS spectra were reduced using the spec2d DEIMOS data reduction pipeline \citep{Newman2013} adapted by \citet{Kirby2015}. Each spectrum was flat-fielded, wavelength-calibrated, sky-subtracted, and telluric-corrected. We did not flux-calibrate the spectra, as we are interested in continuum-normalized spectra.

We selected the final sample in a manner similar to Paper I with the following criteria. First, the member galaxies are quiescent, which we defined as having equivalent widths (EWs) of [\ion{O}{2}] $\lambda3727$ less than 5 \angstrom\ and either having rest-frame ${\rm FUV}-V$ colors larger than 3 or no detection in rest-frame FUV. Second, the signal to noise (S/N) is greater than 10 \angstrom$^{-1}$ in rest-frame. This resulted in the final sample of 59 galaxies in Cl0024 and 92 galaxies in MS0451. The lowest masses are $M_*=10^{9.7}$ and $10^{9.6}~M_\odot$, respectively (see Festar
igure \ref{fig:datademo}).

Finally, we include a subsample of $z\sim0$ SDSS quiescent galaxies in our sample to compare with our observed galaxies. We use the same 155 randomly selected SDSS quiescent galaxies from \citeauthor{Gallazzi2005}'s (\citeyear{Gallazzi2005}) sample in the mass range $10^{9}$ to $10^{11.5} M_\odot$ as in Paper I\@. To recapitulate, we put limits on broadband color and on the maximum H$\alpha$ EW so that the sample selection criteria are close to what we used for the higher redshift sample. We select roughly equal numbers of SDSS galaxies in each mass bin.

\section{Model Fitting}
\label{sec:modelfitting}
\begin{figure*}
    \centering
    \includegraphics[width=0.85\textwidth]{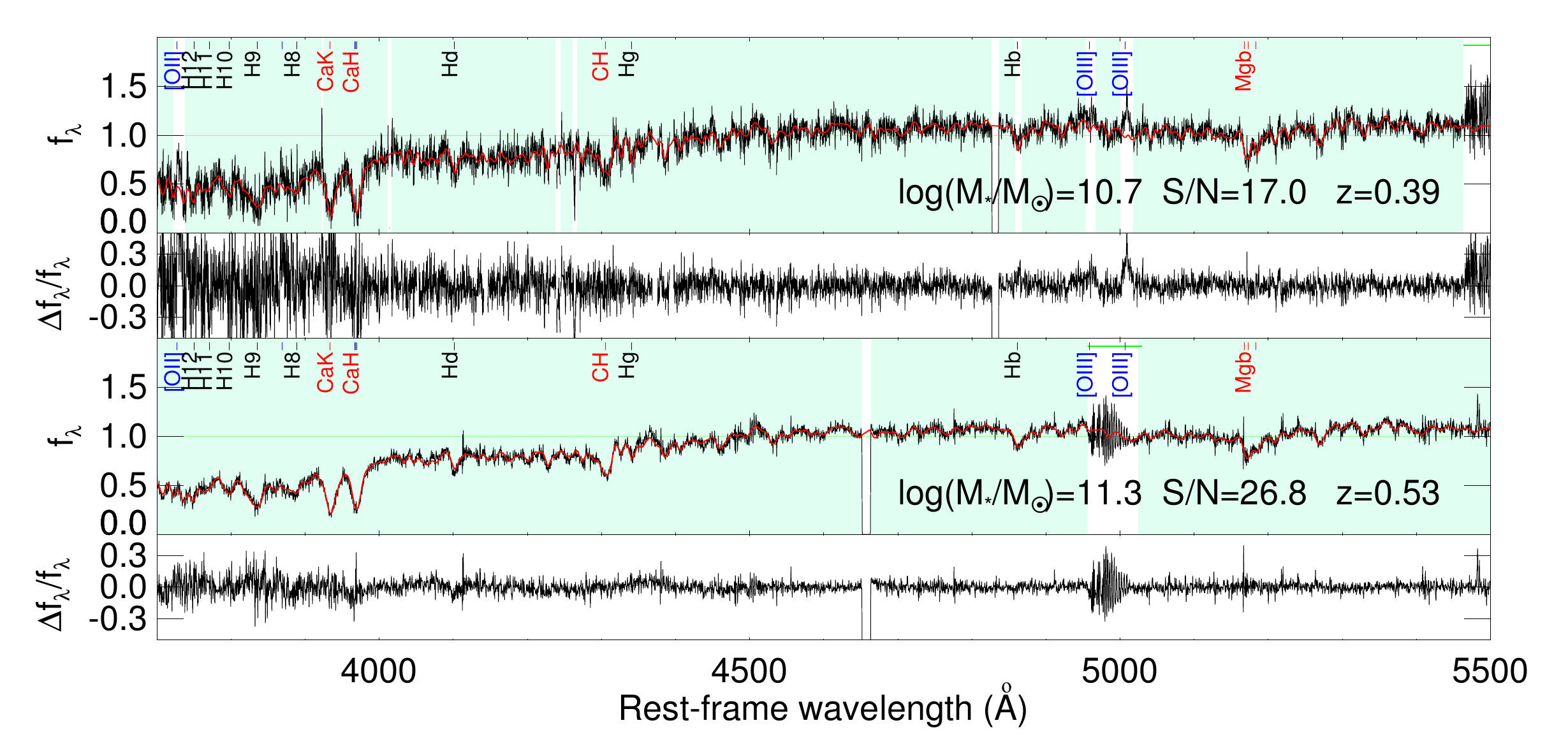}
    \caption{Examples of two observed spectra (black), best fit models (red), and the model residual (bottom panel of each spectrum). The teal background shows the spectral regions used for spectrum modeling while the white background shows the spectral regions that are masked out. The top panels are an example of the spectra at the lower end of the S/N cut from a member of Cl0024 cluster. The bottom panels show a high S/N spectrum from the MS0451 cluster. The best-fit parameters are Age$=3.3^{+1.4}_{-0.7}$ and $4.3^{+1.7}_{-0.9}$ Gyr; [Fe/H]$=-0.0^{+0.11}_{-0.7}$ and $-0.05^{+0.08}_{-0.06}$ dex; [Mg/Fe]$=0.39^{+0.13}_{-0.11}$ and $0.39^{+0.15}_{-0.04}$ dex, respectively.}
    \label{fig:examplespec}
\end{figure*}

 We use full spectral fitting to derive ages and stellar metallicities. We develop our own fitting algorithm in this work, although spectral fitting algorithms capable of deriving stellar population properties of stellar population already exist. Recent examples of existing algorithms include FIREFLY \citep{Wilkinson2017} and \texttt{alf} \citep{Conroy2018}. We use our own code because we are interested in the abundances of specific elements. \texttt{alf} is the only algorithm that is currently readily capable for this, but the Markov chain Monte Carlo method that it uses to find best-fit parameters is computationally expensive when applied to a large sample size. This is especially pertinent because we wanted to experiment using different spectral masks and continuum normalization techniques. Lastly, we demonstrate in Appendix \ref{sec:testcomb} that our test results agree reasonably well (within $2\sigma$) with those obtained using \texttt{alf}.
 
 As in \citet{Leethochawalit2018}, we adopt the single stellar population (SSP) models from the Flexible Stellar Population Synthesis \citep[FSPS,][]{Conroy2009} version 3.0. We generate the SSP spectra with the \citet{Kroupa2001} IMF, Padova isochrones \citep{Marigo2007}, and the MILES spectral library \citep{Sanchez2006}. The models have metallicity and age ranges of $-1.98<\log Z<0.2$ and 0.3 Myr to 14 Gyr, respectively. Unlike the model used in \citet{Leethochawalit2018}, here we include the enhancements of Mg in addition to [Fe/H]. To do so, we use the theoretical response functions from \citet{Conroy2018}, which depend on metellicity and age. They were computed from the Kurucz suite of theoretical model atmospheres and spectra \citep{Kurucz1993}. 

We are interested in 3 parameters: age, [Fe/H], and [Mg/Fe]. We choose to measure [Mg/Fe] for two reasons. First, magnesium is one of the alpha elements that is mainly produced by Type II supernovae. Its recycling time can thus be approximated as instantaneous. Second, magnesium absorption features are distinct. Specifically, the Mg~b absorption lines at 5170 \angstrom\ minimally overlap with absorption features of other elements. This makes the measurement more reliable than other instantaneously recycled elements, especially in low-S/N spectra.

We assume that each galaxy is a single stellar population, i.e., all stars in the galaxy were born at the same time with the same metallicity. Thus, our obtained ages and metallicities are SSP-equivalent values. This means that the measured age is expected to be younger than both light- and mass-weighted ages derived in models with extended star-formation histories \citep{Choi2014}. The metallicities are less affected \citep{Mentz2016}. 

In order to measure age, [Fe/H], and [Mg/Fe], we actually need to fit for 6 parameters, all of which influence the 3 parameters of interest: age, [Z/H], velocity dispersion, redshift, [Mg/Fe] and [N/Fe]. Although we are mainly interested in [Mg/Fe], we experimented with a few options for the combinations of additional metal enhancements in Appendix \ref{sec:testcomb}. In summary, we find that by fitting for two additional abundance ratios---[Mg/Fe] and [N/Fe]---we obtain the values of interest that are most consistent with literature and with the results from fitting with a more elaborate set of metals: [Mg/Fe], [O/Fe], [C/Fe], [N/Fe], [Na/Fe], [Si/Fe], [Ca/Fe], and [Ti/Fe]. This is likely because the flux of the wavelength within the range of $\sim4000-4400$ \angstrom\ responds simultaneously to the enhancements of Fe, N, and Mg.

We interpret the measured [Z/H] as [Fe/H] for the following reason. In our case, the measured [Z/H] is the metallicity of the base SSP model (without Mg and N enhancement) and therefore not the actual total metallicity of the galaxy. Each base SSP model is the integrated stellar spectrum of stars that lie on the isochrone of a given [Z/H] and age. However, the Padova isochrones assume solar-scaled abundances of individual metal elements (${\rm [Fe/H]}={\rm [Z/H]}$). In addition, the metallicities in the MILES stellar library were measured in terms of [Fe/H]. Thus, our measured [Z/H] is best interpreted as [Fe/H]. Besides, based on results in Appendix \ref{sec:testcomb}, our measured [Fe/H] agrees well with [Fe/H] measured from more complex models.

The fitting method is the same as that in Paper I. In summary, we first mask out the most prominent telluric band in the 7591--7703 \angstrom\ wavelength range. In addition, if the spectrum has a positive EW of the [\ion{O}{2}]$\lambda3727$ emission line, we then also mask out the wavelength range of the [\ion{O}{2}]$\lambda3727$, [\ion{O}{3}]$\lambda5007,4959$ and H$\beta$ line. We iteratively fit each spectrum with the Levenberg--Marquardt $\chi^2$-minimization method. 

For each spectrum, we fit for at least 100 iterations with the IDL code MPFIT \citep{Markwardt2012} until the fitting parameters converge. In the first iteration, we fit the continuum-normalized observed spectrum with continuum-normalized model spectra. 
In the subsequent iterations, we do not alter or continuum-normalize the model spectra. Instead, we apply a synthesized continuum (a B-spline fit to the quotient of the continuum-normalized spectrum and the best-fit SSP model spectrum from the previous iteration) to the observed continuum-normalized spectrum. We then fit the resulting spectrum with unaltered model spectra. This algorithm avoids having to separately determine the continuum for the observed and model spectra. The method was described in full detail in Paper I. Lastly, to avoid the convergence of parameters on local minima, we fit each spectrum at least 5 times with different initial parameters, and we adopt the results from the models with the least $\chi^2$. 

Finally, we convert the measured enhancement of the [Mg/Fe] from the base FSPS spectra to the actual [Mg/Fe]. We followed the method in \citet{Conroy2018} by adding the abundance pattern of the MILES library stars at a given [Fe/H] to the measured Mg enhancement to obtain the final [Mg/Fe]. Because the [Mg/Fe] pattern of the stellar library in the measured [Fe/H] range only varies within $\sim0.05$ dex, this step does not significantly affect the reported value of [Mg/Fe]. 

As demonstrated in Paper I, the statistical uncertainties obtained from MPFIT underestimate the total uncertainties. Likewise, here we also calculate the systematic uncertainties based on the degeneracies between the parameters of interest and take those as our uncertainties. We first generate a mock spectrum with the measured [Fe/H], [Mg/Fe], and age. We then smooth the spectrum to a fixed velocity dispersion of 250 km/s, convolve it with the observed spectral resolution, and add Gaussian noise to reach the same S/N of the observed spectrum. We thereafter compare the noised spectrum to a 3-dimensional grid of noise-less SSPs (in age, [Fe/H] and [Mg/Fe]), 
and calculate the uncertainties based on the $\chi^2$ grid.\footnote{Each noise-less SSP grid contains $51\times51\times41$ spectra, covering the range of $[-1.0,2.5]$ dex in [Fe/H], [0.1,12] Gyr in age, and [-0.4,0.8] dex in [Mg/Fe], respectively. The reason for fixing the velocity dispersion at 250 km/s is therefore to keep the size of the SSP grids manageable.}  

We explore how the uncertainties of the three parameters correlate with each other at a given S/N ratio in Appendix \ref{appendix:uncertainties}. In summary, the uncertainties in [Fe/H] and age mostly originate from the highly-correlated degeneracy between age and [Fe/H]. There is also a positive correlation between the uncertainties in [Fe/H] and in [Mg/Fe]. These result in average uncertainties of $\sim0.2$ dex at S/N$=10$\angstrom$^{-1}$ and $\sim0.1$ dex at S/N$=25$\angstrom$^{-1}$ in age and [Fe/H]; both can be higher at lower [Fe/H] values. The uncertainties of [Mg/Fe] are $\sim0.2-0.3$ at S/N$=10$\angstrom$^{-1}$, and $\sim0.1-0.2$ dex at 25 \angstrom$^{-1}$.

Examples of best-fit models of the spectra at the lower and upper end of the S/N range are shown in Figure \ref{fig:examplespec}. We list all measured parameters in Table \ref{table:allpar}.

\begin{deluxetable*}{lllCCCCC}
\tablewidth{0pt}
\tablecolumns{8}
\tablecaption{Catalog of Measured Age and Metalicities\label{tab:catalog}}
\tablehead{\colhead{No.} & \colhead{RA} & \colhead{DEC} & \colhead{$\log(M_*/M_\odot)$}& \colhead{Age (Gyr)} & \colhead{[Fe/H]}  & \colhead{[Mg/Fe]} & \colhead{$\eta$}}
\startdata
  \multicolumn{8}{c}{Cl0024+17}\\
  1 & 00 25 51.07 & +17 08 42.4 &  10.8 &  $  2.2^{  +0.4}_{  -0.2}$ &  $+0.19^{+0.09}_{-0.13}$ &  $+0.02^{+0.15}_{-0.13}$ & $0.46^{+0.83}_{-0.17}$\\
  2 & 00 25 54.52 & +17 16 26.4 &  10.4 &  $  3.0^{  +1.9}_{  -1.1}$ &  $-0.30^{+0.24}_{-0.16}$ &  $+0.25^{+0.20}_{-0.22}$ & $1.25^{+1.97}_{-0.78}$\\
  3 & 00 25 57.73 & +17 08 01.5 &  10.4 &  $  3.5^{  +2.4}_{  -1.0}$ &  $+0.06^{+0.12}_{-0.09}$ &  $+0.04^{+0.16}_{-0.18}$ & $0.75^{+1.11}_{-0.36}$\\
  4 & 00 26 04.30 & +17 18 46.5 &  10.3 &  $  2.2^{  +0.9}_{  -0.7}$ &  $-0.42^{+0.23}_{-0.14}$ &  $+0.08^{+0.32}_{-0.30}$ & $3.00^{+3.66}_{-2.07}$\\
  5 & 00 26 04.44 & +17 20 00.6 &  10.5 &  $  2.9^{  +1.6}_{  -1.4}$ &  $-0.21^{+0.26}_{-0.14}$ &  $+0.56^{+0.21}_{-0.22}$ & $0.19^{+0.77}_{-0.19}$\\
  \multicolumn{8}{c}{...}\\
  \multicolumn{8}{c}{MS0451}\\
  1 & 04 53 18.17 & -02 58 57.7 &  11.1 &  $  3.4^{  +2.6}_{  -1.0}$ &  $+0.05^{+0.13}_{-0.09}$ &  $+0.56^{+0.16}_{-0.15}$ & $\textrm{N/A}$\\
  2 & 04 53 33.42 & -02 56 23.5 &  10.9 &  $  1.9^{  +0.3}_{  -0.3}$ &  $-0.07^{+0.06}_{-0.12}$ &  $+0.49^{+0.16}_{-0.15}$ & $0.08^{+0.56}_{-0.08}$\\
  3 & 04 53 36.54 & -03 04 13.5 &  11.3 &  $  4.8^{  +2.0}_{  -1.4}$ &  $-0.06^{+0.09}_{-0.06}$ &  $+0.50^{+0.11}_{-0.11}$ & $0.06^{+0.40}_{-0.06}$\\
  4 & 04 53 38.64 & -02 54 11.2 &  10.8 &  $  3.0^{  +1.4}_{  -0.7}$ &  $-0.08^{+0.21}_{-0.13}$ &  $+0.13^{+0.15}_{-0.18}$ & $0.92^{+1.36}_{-0.59}$\\
  5 & 04 53 52.61 & -02 55 31.5 &  10.8 &  $  1.9^{  +0.4}_{  -0.3}$ &  $-0.35^{+0.08}_{-0.23}$ &  $+0.42^{+0.20}_{-0.16}$ & $0.84^{+1.35}_{-0.22}$\\
 \multicolumn{8}{c}{...}
 \enddata
 \tablecomments{Full table is available online in machine-readable form. Note that the [Fe/H]'s listed here are different from those in Paper I, which were calculated by masking out the Mg b lines and assuming the solar abundance ratios. See Appendix \ref{sec:testcomb} for the comparison between the two results. The mass-loading factors, $\eta$, are calculated based on Equation \ref{equation:quiescentz} using the fiducial yield and return fraction. Where $\eta$ is entered as N/A, the [Mg/H] is larger than the adopted yield - see Section \ref{section:massloading1} for a more detailed discussion. }
 \label{table:allpar}
\end{deluxetable*}

\section{Results}\label{sec:results}
\subsection{The MZR measured from [Fe/H] evolves with redshift}
\begin{figure*}
   \hspace{-0.3in}   
   \includegraphics[width=0.54\textwidth]{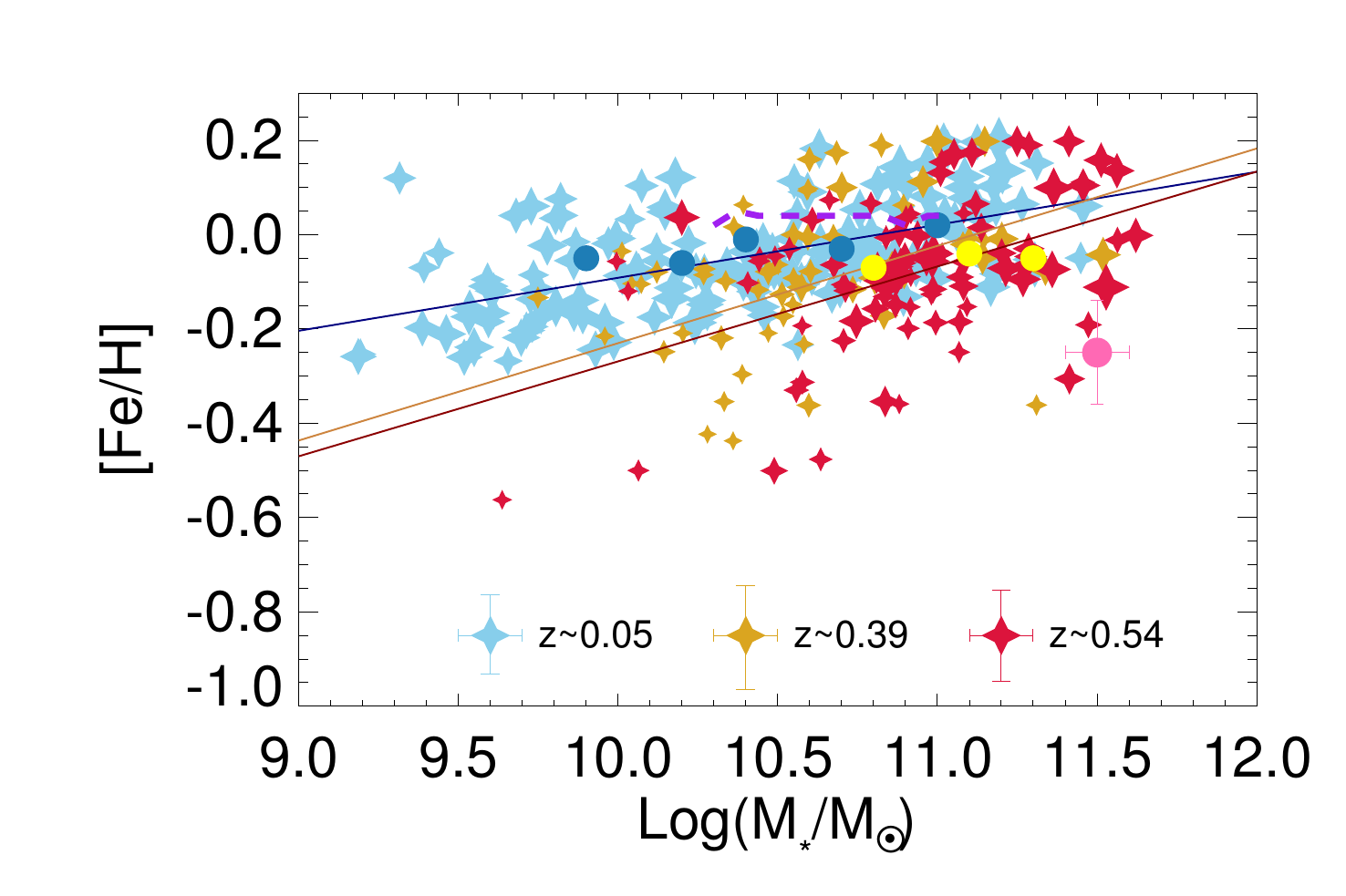}   
   \hspace{-0.3in}
   \includegraphics[width=0.54\textwidth]{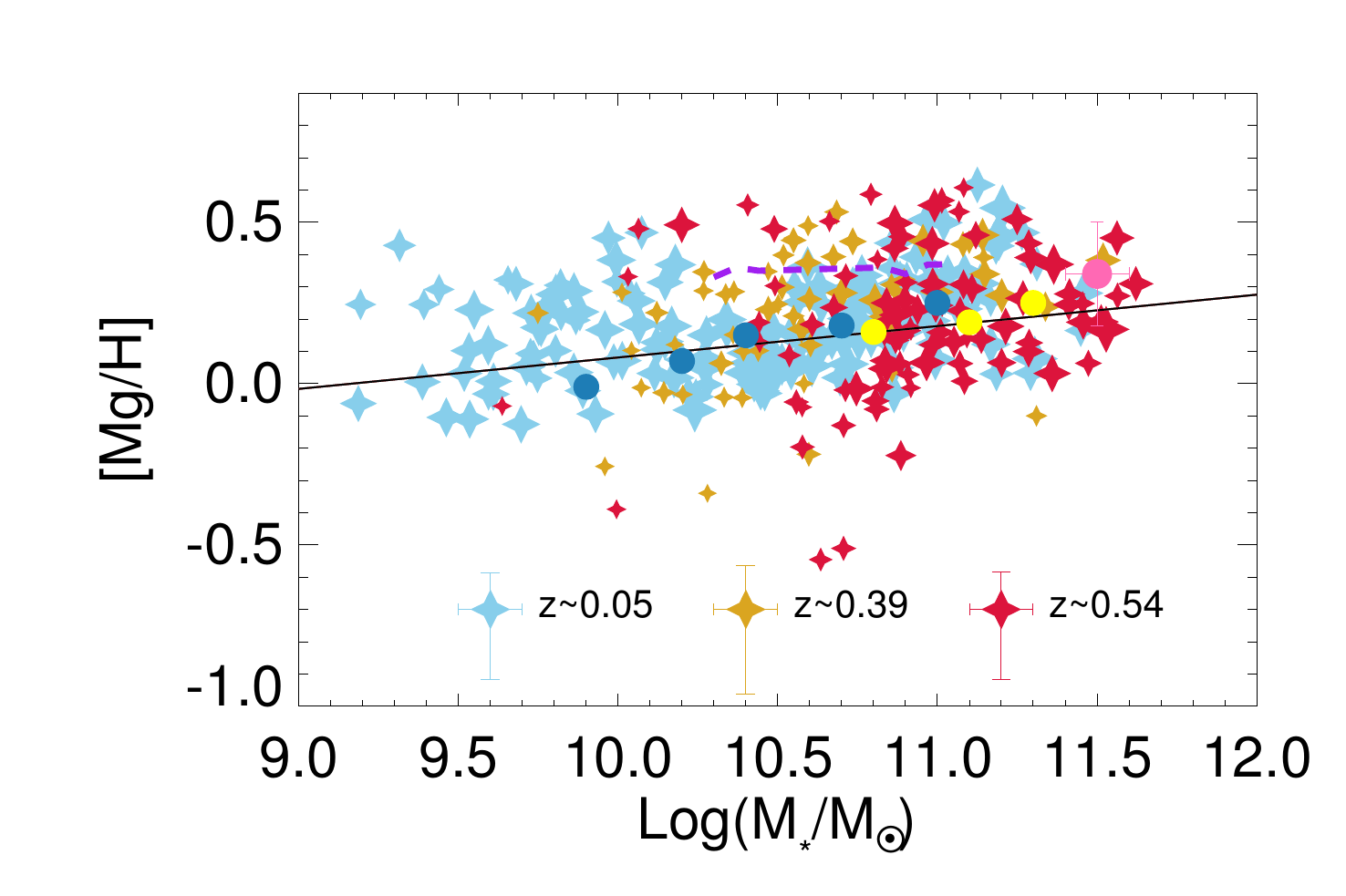}
   \caption{Stellar mass--metallicity relations measured from [Fe/H] (left) and [Mg/H] (right). Measurements in this paper are in diamonds. The labels indicate different redshift samples and their median uncertainties. The size of the diamonds corresponds to the S/N of the observed spectra, ranging from 10 \angstrom$^{-1}$ (smallest) to 45 \angstrom$^{-1}$ (largest). Best-fit linear models from Table \ref{table:feh_linearparam} are plotted in solid lines (model 1 for [Fe/H] and model 3 for [Mg/H]). Measurements based on stacked spectra of similar redshifts by \citet{Choi2014} are in dark blue and yellow circles (for $z=0.1$--$0.2$ and $z=0.4$--$0.55$ redshift bins, respectively). The pink circle represents a measurement of a $z\sim2.1$ quiescent galaxy from \citet{Kriek2016}. Magenta dashed lines are predictions based on the $z\sim0$ quiescent galaxies with $10^{10.2}\lesssim M_*/M_\odot\lesssim10^{11}$ in the IllustrisTNG simulation \citep{Naiman2018}. We found a significant evolution with redshift in the [Fe/H]--mass relation, but not in the [Mg/H]--mass relation.\label{fig:allmzr}}
\end{figure*}

We show the relation between [Fe/H] and galaxy stellar mass in Figure \ref{fig:allmzr}. Visually, the galaxies at higher redshift (Cl0024 and MS0451) appear to have lower [Fe/H] abundances than those of local SDSS galaxies, especially in the lower mass range. A small proportion of data points appear to have very low [Fe/H] or [Mg/H] values. These are mostly low S/Ns spectra and younger populations. In the left panel, there are 17 data points whose [Fe/H]'s are smaller than 0.15 dex below the best-fit linear relations. Nevertheless, only eight of them actually have [Fe/H]'s that are inconsistent with the best-fit lines within $2\sigma$ of uncertainty. These galaxies are mostly younger than 2 Gyrs, which is the age range that is prone to increased bias \citep[e.g.,][]{Ge2018,CidFernandes2018}. As for the right panel, there are 5 galaxies whose $2\sigma$ upper limits of [Mg/H] are not consistent with the best-fit line (but all are consistent within $3\sigma$). Interestingly, these galaxies do not overlap with the galaxies with very low [Fe/H]. 

Next, we use the analysis of covariance to quantify the dependence of MZR on redshift. We first assume that both normalization and slope of the mass--[Fe/H] relation depend on redshift. We fit data from every redshift at once with a single linear multiple regression, where [Fe/H] depends on one quantitative variable (mass) and categorical variables (different redshift samples). We choose to fit with a linear function because we found in Paper I that both a two-degree polynomial and a logarithmic function introduce degrees of freedom that are not justified by the data.

The first model we consider (model 1), allows interactions\footnote{``Interaction" is a term in statistics that describes a situation where the effect of one independent variable depends on the magnitude of the other independent variable. In this case, we allow the evolution of metallicity with redshift to be mass-dependent.} between mass and redshift. This means that \textit{normalizations} and \textit{slopes} of the MZRs are allowed to depend on redshift. The model is
\begin{equation}
\begin{split}
    E&(\text{[Fe/H]}\mid M_{10},\text{sample})=  \beta_0+\beta_{M}M_{10}+\beta_{cl}cl\\
    & +\beta_{ms}ms +\beta_{cl\cdot M}(cl\cdot M_{10})+\beta_{ms\cdot M}(ms\cdot M_{10})
\end{split}
\label{equation:full_lineq}
\end{equation}
$cl$ (or $ms$) is an indicator or a categorical variable with a value of 1 if the data point belongs to the Cl0024 (or MS0451) sample, and 0 otherwise. $M_{10}$ is the galaxy stellar mass, defined as $M_{10} = \log[M_*/10^{10}M_\odot]$ so that the normalization is at $10^{10} M_\odot$. To digest this equation, we can view that $\beta_0$ is the normalization and $\beta_{M}$ is the slope of the MZR for the $z\sim0$ SDSS population; we call this the baseline MZR. The rest of the parameters are the evolution terms. $\beta_{cl}$ is the change in the normalization, while $\beta_{cl\cdot M}$ is the change in the slope from those of local galaxies to those of $z\sim0.4$ Cl0024 galaxies. $\beta_{ms}$ and $\beta_{ms\cdot M}$ can be interpreted similarly.

We use the Monte Carlo (MC) technique with 1000 iterations to find the best-fit $\beta_i$ parameters. In each iteration, we draw all samples randomly according to the measured probability distribution of [Fe/H]. We use the IDL code \texttt{regress} to find the best-fit parameters from each iteration. The final best-fit parameters (the mean and standard deviation) are listed as model 1 in Table \ref{table:feh_linearparam}.

\begin{deluxetable*}{cllrlrrlr}
\tablecaption{Model parameters for MZRs\label{table:feh_linearparam}}
 \tablehead{
 \colhead{} &\colhead{} &\colhead{}  & \multicolumn{3}{c}{[Fe/H] vs mass} & \multicolumn{3}{c}{[Mg/H] vs mass} \\
 \colhead{model} & \multicolumn{2}{l}{parameter/description}&\colhead{value} &  \colhead{t} & \colhead{sig.}&\colhead{value} &  \colhead{t} & \colhead{sig.}}
\startdata
1 &$\beta_0$     & constant    & $-0.09 \pm\ 0.01$ & 7.9 & $<$0.0005 & $0.09 \pm 0.02$ & 4.5 & $<0.0005$\\
  &$\beta_{M}$   & mass        & $0.11 \pm\ 0.02$ & 7.2 & $<$0.0005 & $0.10 \pm 0.03$ & 3.8 & $<0.0005$\\
  &$\beta_{cl}$  & cl0024 & $-0.14 \pm\ 0.05$ & 2.9 & 0.002 & $-0.06 \pm 0.07$ & 0.8 & 0.2\\
  &$\beta_{ms}$  & ms0451 & $-0.18 \pm\ 0.05$ & 3.9 & $<$0.0005 & $-0.01 \pm 0.07$ & 0.2 & 0.4\\
  &$\beta_{cl\cdot M}$& cl0024*mass & $0.09 \pm\ 0.06$ & 1.6 & 0.05 & $0.12 \pm 0.09$ & 1.4 & 0.08\\
  &$\beta_{ms\cdot M}$& ms0451*mass & $0.09 \pm\ 0.04$ & 2.0 & 0.02 & $-0.03 \pm 0.07$ & 0.4 & 0.4\\
\hline
2 &$\beta_0$  & constant& $-0.10 \pm\ 0.01$ & 9.3 & $<0.0005$ & $0.09 \pm 0.02$ & 4.2 & $<0.0005$\\
  &$\beta_{M}$& mass    & $0.14 \pm\ 0.02$ & 9.3 & $<0.0005$ & $0.11 \pm 0.02$ & 4.5 & $<0.0005$\\
  &$\beta_{cl}$& cl0024 & $-0.09 \pm\ 0.02$ & 3.9 & $<0.0005$ & $0.02 \pm 0.03$ & 0.5 & 0.6\\
  &$\beta_{ms}$& ms0451 & $-0.11 \pm\ 0.02$ & 6.4 & $<0.0005$ & $-0.04 \pm 0.03$ & 1.4 & 0.2\\ 
\hline
3 &$\beta_0$  & constant& & & & $0.08 \pm 0.02$ & 4.2 & $<0.0005$\\
  &$\beta_{M}$& mass    & & & & $0.10 \pm 0.02$ & 4.4 & $<0.0005$\\
\hline
\multicolumn{3}{l}{Is model 1 necesary over model 2?} & \multicolumn{2}{c}{likely} & 0.01 & \multicolumn{2}{c}{no} & 0.3\\
\multicolumn{3}{l}{Is model 2 necessary over model 3?} & & &  & \multicolumn{2}{c}{no} & 0.2
\enddata
\tablecomments{The significance values (sig.) are the p-values from one-tailed t tests. The values will be twice for the two-tailed test. We use model 1 for [Fe/H] and model 3 for [Mg/H] relations with mass for further discussion.}
\end{deluxetable*}

Based on model 1, the MZRs of the higher redshift samples are likely steeper than that of local galaxies. The $\beta_{cl\cdot M}$ and $\beta_{ms\cdot M}$ terms suggest that the mean change in [Fe/H] for a 1 dex increase in stellar mass is $\sim0.9$ dex more for higher redshift galaxies than for local galaxies. However, their significance values are only 2--3$\sigma$. 

We further formally test the necessity of using the interaction model (model 1) over a simpler model (model 2) without the interactive terms, i.e., the model where the slopes are fixed to the same value regardless of redshift:  
\begin{equation}
    E(\text{[Fe/H]}\mid M_{10},\text{sample})=  \beta_0+\beta_{M}M_{10}+\beta_{cl}cl+\beta_{ms}ms 
\end{equation}
The best-fit parameters are also listed in Table \ref{table:feh_linearparam}. The F-test for the $R^2$ change between the two models suggests that we can reject the null hypothesis that the simpler model 2 is sufficient to describe the data with a $p$-value of 0.01 (also $>2\sigma$). Despite these borderline values of 2--3$\sigma$, for [Fe/H], we conclude that the interaction model (model 1) performs better than the simpler model 2 and we will work only with model 1 for further interpretations.

We found that the normalizations for the MZRs of the two higher redshift samples are statistically significantly lower ($>3\sigma$) than the normalization of the local SDSS galaxies. With the more detailed spectral fitting model and larger sample size in this work, we confirm the finding in Paper I that there is an evolution in the MZR when the metal indicator is [Fe/H]. The best estimate of normalizations at $10^{10} M_\odot$ for the $z\sim0.39$ and the $z\sim0.54$ samples are $\sim0.14$ and $0.18\pm0.05$ dex lower than the local samples. These values translate to the evolution of $0.04\pm0.01$ dex per observed Gyr, consistent with what we reported in Paper I. 

In conclusion, we found that model 1 is the most appropriate to describe [Fe/H] evolution. We observe an evolution in [Fe/H] in both slope (likely at $2-3\sigma$) and normalization ($>3\sigma$). This means that the amount of the evolution with observed redshift depends on mass. Over the same redshift range, lower mass quiescent galaxies evolve more strongly in [Fe/H] at fixed mass than higher mass quiescent galaxies. Further interpretation of [Fe/H] in Section \ref{sec:discussion} will be mainly based on the results from this model.

\subsection{The MZR measured from [Mg/H] does not evolve with observed redshift}
Now, we look at the evolution of the relation between [Mg/H] and $M_*$. First, we repeat the fitting with model 1 but substitute [Fe/H] with [Mg/H]. We found that the mass--[Mg/H] relations of each redshift sample have the same slope within the $\sim 1\sigma$ uncertainties. The best-fit MZRs at different redshifts are roughly parallel. This suggests that the simpler model 2, in which slopes are fixed to a common value, is more appropriate to describe the mass--[Mg/H] relation than the more complex model 1. 

We continue to fit with model 2 and also found no significant evolution in the normalization of the MZR with observed redshift. As shown in Table \ref{table:feh_linearparam}, the values of $\beta_{cl}$ and $\beta_{ms}$ in model 2 are both consistent with zero. We found no significant evolution in [Mg/H] with observed redshift.

The results from the first two models suggest that we should proceed to fit a simple linear equation without redshift-dependent parameters (model 3). The parameters are also listed in Table \ref{table:feh_linearparam}. This model is most appropriate for the mass--[Mg/H] relation. 

In summary, we did not detect an evolution of the stellar mass--[Mg/H] relation with redshift, neither in terms of slope nor normalization. This is in contrast to the $>3\sigma$ detected in the evolution of the stellar mass--[Fe/H] relation with redshift. We can explicitly separate the best-fit parameters for both Fe and Mg abundances into the predicted MZRs at each redshift. The MZRs based on the measurements in this work are the following: 
\begin{equation}
\small
\begin{split}
   \text{[Fe/H]}(M_{10},z\sim0)&= (-0.09\pm0.01)+(0.11\pm0.02)M_{10}\\
    \text{[Fe/H]}(M_{10},z\sim0.39)&= (-0.23\pm0.05)+(0.20\pm0.06)M_{10}\\
    \text{[Fe/H]}(M_{10},z\sim0.54)&= (-0.27\pm0.05)+(0.20\pm0.04)M_{10}\\
    \text{[Mg/H]}(M_{10}) &= (+0.08\pm0.02)+(0.10\pm0.02)M_{10}\\
\end{split}
\label{eq:explicit_mzr}
\end{equation}
where $M_{10} = \log[M_*/10^{10}M_\odot]$.  
\subsection{Intrinsic Scatter}
We measured the intrinsic scatter ($\sigma_Z$) in the [Fe/H] and [Mg/H] relations with mass (Figure \ref{fig:allmzr}).  We used a method similar to that in Paper I\@. In short, we convolved a normalized Gaussian with size $\sigma_Z$ to the existing probability of each data point. The measured intrinsic scatter is the $\sigma_Z$ that maximizes the sum of the resulting log-likelihood evaluated at the predicted [Fe/H] (or [Mg/H]) according to the relations in Equation \ref{eq:explicit_mzr}. The uncertainties of the intrinsic scatter are derived via the jackknife resampling technique.

The measured intrinsic scatter in [Fe/H] is consistent with what we found in Paper I: $0.06\pm0.01$ dex, compared to $0.07\pm0.01$ dex in Paper I\@. The nominal intrinsic scatter in [Mg/H] is smaller, but it has a large uncertainty: $0.05\pm0.17$ dex, which is consistent with both zero and with the intrinsic scatter of [Fe/H]. As a test, if we remove the two data points whose [Mg/H] is below $-0.4$ dex, the measured $\sigma_Z$ in [Mg/H] becomes $0.03\pm0.05$ dex, which is still consistent with both zero and the intrinsic scatter of [Fe/H]. If we expect that the true intrinsic scatter is non-zero, our result, by being consistent with zero, indicates that we may have overestimated the uncertainties in [Mg/H]. In addition, if the true intrinsic scatter in [Mg/H] is small (e.g., less than 0.08 dex), we may have a small number of ``outliers'' in the data for which we underestimated the uncertainties.

In sum, our results indicate that the intrinsic scatter in [Mg/H] is likely smaller than the intrinsic scatter in [Fe/H]. However, we refrain from any further interpretation due to the large uncertainties in our measurements. 

\section{Discussion}\label{sec:discussion}
In this section, we will interpret the measured MZRs and their evolution with redshift, or lack thereof, using galactic chemical evolution models. Specifically, we use the measured MZRs to constrain average galactic outflows in terms of the mass-loading factor.

\subsection{Comparison to galaxy simulations and semi-analytic models}
The relatively shallow slopes of the MZRs estimated from our samples are consistent with other observations that specifically sample quiescent galaxies. Our slopes for both [Fe/H] and [Mg/H] are consistent within $2\sigma$ with the slopes derived from simple linear fits to the measurements by \citet{Choi2014}, which were based on stacked spectra of field quiescent galaxies at similar redshifts. They are also consistent with the slopes for the quiescent galaxies reported by \citet{Gallazzi2014}, in which metallicities ($Z$) were measured from a combination of H, Fe, and Mg absorption lines. 

\citet{Gallazzi2014} found that the slopes of the MZRs of quiescent galaxies are shallower than those of star-forming galaxies.\footnote{This is however not true for Local Group dwarf galaxies \citep{Kirby2013}.} Do theoretical models reproduce the shallower slope?

Several semi-analytical (SA) and hydrodynamical simulation models have predicted stellar MZRs and their evolution with redshift \citep[e.g.][]{Lu2014, Guo2016, Ma2016, TaylorKobayashi2016}. However, most results from simulations do not report metallicities based on galaxy star formation properties. The predictions are for the total populations, which are mixtures of star-forming and quiescent galaxies, depending on the passive fraction at each redshift.

It is therefore important to be cautious of interpreting the differences between observations and simulations. In Paper I, we compared the slope of [Fe/H] measured in quiescent galaxies to the slopes predicted in SA models in \citet{Lu2014} at face value. We argued that the observational results agree the most with the model in which the outflow mass-loading factor from Type II SNe is independent of mass. However, the galaxies simulated by \citeauthor{Lu2014} have gas fractions on average ranging from 20\% in the highest mass bin to 50\% in the lowest mass bin at $z\sim0$. In other words, portions of the simulated galaxies are still forming stars, which is not the case for our observed samples.

One study that specifically predicts the MZRs of quiescent galaxies is by \citet{Naiman2018}, using the IllustrisTNG suite of simulations. The median of their simulated stellar abundances at $z\sim0$ is plotted as magenta lines in Figure \ref{fig:allmzr}. We calculated the simulated [Mg/H] based on their median [Fe/H] and [Mg/Fe]. The slope in the simulated [Fe/H] is flatter than the observations. The normalization is consistent with our observations at $\sim10^{11}M_\odot$ but is larger at lower masses. The relatively flat slope in the simulations was already explained by \citeauthor{Naiman2018} as a result of the sampling criteria. Specifically, they excluded smaller mass galaxies in their sample. If those were included, they reported that the predicted mass--[Fe/H] relation could have been steeper. For [Mg/H], the simulated slope is $\sim0.04$ dex per log(mass), which is roughly half of what we observed. Because we calculated [Mg/H] from [Mg/Fe] and [Fe/H], it is likely that the cause of the difference in [Mg/H] slopes between the observations and simulations is the same as for [Fe/H].

\subsection{Analytic chemical evolution model}
As an alternative to more complex simulations, we can schematically understand the stellar metallicities of quiescent galaxies using an analytic galactic chemical evolution model. In this paper, we present a simple model that connects the metal mass in a quiescent galaxy to the average outflow it experienced. The model is based on the work of \citet{Lu2015}. In this model, the assumptions are

\begin{enumerate}
\item Star formation occurs in the interstellar medium (ISM) that is perfectly mixed.
\item Metals are instantaneously recycled.
\item Outflows and inflows are permitted but the inflow do not contribute significantly to the total metal budget.\footnote{We also derive an estimate for stellar metallicity when the inflow is enriched and when the outflow has different metallicity than the ISM, as well as evaluate these effects on our results in Appendix \ref{appendix:enrichedoutflow}.}
\end{enumerate}

We start by tracking the change in the total metal mass in the ISM ($dM_{z,g}$) as follows:
\begin{equation}
    dM_{Z,g} = ydM_*-Z_gdM_*-\frac{\eta}{1-R}Z_gdM_*
\label{equation:chem1}
\end{equation}
where $M_*$ is the mass of long-lived stars. The first term on the right side of Equation \ref{equation:chem1} represents the metal mass that is newly produced and returned to the ISM by a generation of forming stars. $y$ is the chemical yield, defined as the mass of metals returned to the ISM per mass turned into low-mass stars and remnants ($dM_*$). The second term is the metal mass in the ISM that is locked into stars. The last term is the metal mass that is lost to outflows, which is equal to the mass outflow rate times gas-phase metallicity $Z_g$. The outflow rate is parameterized in terms of the mass-loading factor ($\eta$), defined as the ratio of the mass outflow rate to the star formation rate (SFR). In this equation, the SFR is written in terms of $dM_*$ and the return mass fraction $R$, which is the fraction of the mass of a stellar generation that returns---with its original composition---to the ISM from short-lived stars and stellar winds.    

With the assumption that the ISM is well mixed, the change in the total metal mass locked in long-lived stars during the time interval in which a mass $dM_*$ of long-lived stars formed is $dM_{Z,*} = Z_gdM_*$. We can substitute this to the last two terms of Equation \ref{equation:chem1} to get  
\begin{equation}
     dM_{Z,g} = ydM_*-(1+\frac{\eta}{1-R})dM_{Z,*}
\label{equation:chem2}     
\end{equation}

We then integrate this equation, divide both sides by $M_*$, and rearrange the terms to get a description for the stellar metallicity of a galaxy:
\begin{equation}
    Z_* = \frac{y-r_gZ_g}{1+\frac{\langle\eta\rangle}{1-R}}
\label{equation:stellarz}
\end{equation}
$r_g$ is the cold gas fraction $M_g/M_*$. We adopt the definition of gas- and stellar-phase metallicity as $Z_g=M_{Z,g}/M_g$ and $Z_*=M_{Z,*}/M_*$.
Therefore, in this equation, $Z_g$ is the current gas-phase metallicity and $Z_*$ is already an average metallicity, specifically the mass-weighted metallicity of the galaxy. Note that during the integration, we consider $y$ and $R$ as constant. $\langle\eta\rangle$ can be viewed as the average mass-loading factor weighted by $dM_{Z,*}$, which means that it is most sensitive to feedback at the galaxy's peak of star formation.

Now, we consider stellar metallicities of quiescent galaxies based on Equation \ref{equation:stellarz}. We can see that at the limit where there is no gas left ($r_g=0$), the second term in the numerator is zero and the stellar metallicity is a function of the mass-loading factor.  This situation is appropriate to describe most quiescent galaxies.

We estimate that the product of gas fraction and gas-phase metallicity ($r_g Z_g$) in quiescent galaxies is at most 10\% of the yield. First, quiescent galaxies have low gas fractions. In the local universe, early-type galaxies typically have molecular gas fractions less than a percent \citep{Boselli2014}. 
Galaxies at higher redshifts have higher molecular gas fractions. However, even at $z\sim1.8$, an average early-type galaxy has a gas fraction that is less than $10\%$ \citep{Gobat2018}. Second, quiescent galaxies do not have particularly high gas-phase metallicities. \citet{Griffith2019} measured gas-phase metallicities in 3 quiescent galaxies and found that their $Z_g$ is roughly consistent with their stellar metallicities. This suggests that we can assume quiescent galaxies to have gas fractions $r_g$ less than 10\% and gas-phase metallicities less than the yield, $y$. Therefore, the $r_g Z_g$ term for quiescent galaxies in Equation \ref{equation:stellarz} is at least 1 dex smaller than the yield term.

Considering the discussion above, we estimate the stellar metallicities of quiescent galaxies to be
\begin{equation}
    Z_{*,\text{quiescent}} \approx \frac{y}{1+\frac{\langle\eta\rangle}{1-R}}
\label{equation:quiescentz}
\end{equation}
The yield and return fraction are fundamentally properties of stars. Though they can depend on IMF, they are often treated as constant and not a function of galaxy mass \citep[e.g.][]{Pagel1997}. The only other variable in Equation \ref{equation:quiescentz} is $\langle\eta\rangle$ (hereafter referred to simply as $\eta$).  This means that a non-zero slope of the MZR of quiescent galaxies implies that the mass-loading factor is a function of galaxy mass.

\subsection{Constraints on the mass-loading factor}
\label{section:massloading1}
\begin{figure*}
\centering
    \includegraphics[width=0.7\textwidth]{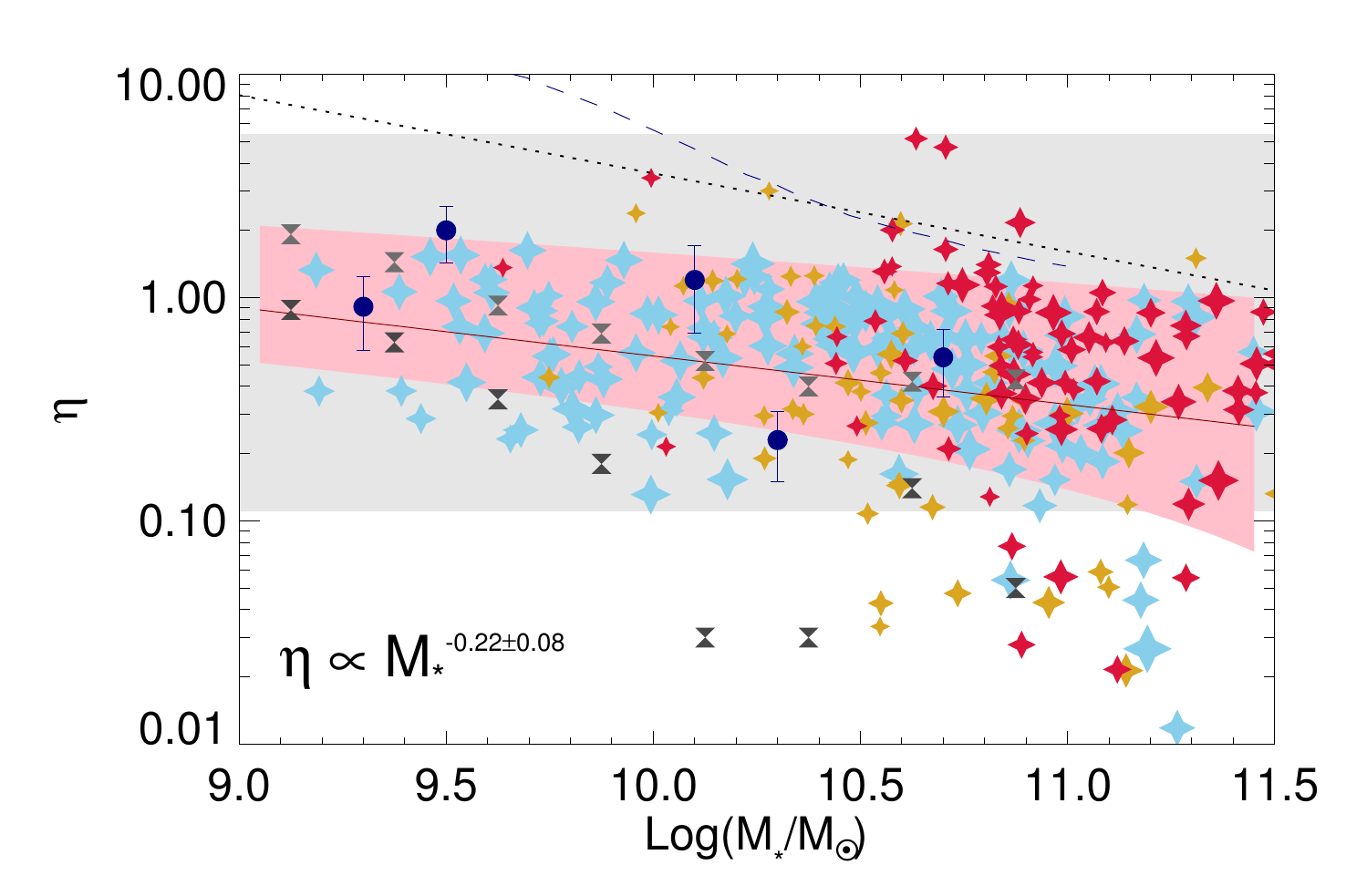}
    \caption{The relation between mass-loading factor ($\eta$) and galaxy stellar mass. Star-shaped symbols show $\eta$ measured from our quiescent galaxy sample, color coded by redshift as in Figure \ref{fig:allmzr}. The median for the uncertainties in the estimated $\eta$ is 0.1 dex. The line is the linear fit to individual measurements of $\eta$. Pink shading represents the best-fit and the scatter in the mass--[Mg/H] relation in Figure \ref{fig:allmzr}. The gray shading and navy dashed line are the estimations based on gas-phase metallicity from \citet{Spitoni2010} and \citet{Lu2015} (upper limits), respectively. he gray hourglasses are the estimations from gas-phase metallicity gradients of nearby star-forming galaxies. The light and dark gray indicate the results using different metallicity calibrations \citep{Belfiore2019}. Navy circles are the mass-loading factors measured from UV absorption lines \citep{Chisholm2017}. Results from the hydrodynamical simulation of \citet{Muratov2015} are shown as the dotted line.} 
    \label{fig:eta}
\end{figure*}
We can now convert measured stellar metallicities of quiescent galaxies into mass-loading factors via Equation \ref{equation:quiescentz}. Because our model assumes instantaneous recycling, we apply our model to the observed Mg abundances. Mg is an $\alpha$ element produced in core-collapse supernovae and is appropriate for the assumption of instantaneous recycling. In addition, comparing to other $\alpha$ elements, it tracks oxygen the closest \citep{Conroy2014}.

We estimate the mass ratio $M_{Mg,*}/M_*$ from the measured [Mg/H] using the solar abundance of Mg from \citet{Asplund2009} and the mass number of Mg (24) as the average Mg to H atomic weight in stars. The yield is set to 3 times the solar Mg abundance \citep{Nomoto2006}. Following \citet{Lu2015}, we set the return fraction to $R=0.46$. 

We found that mass-loading factor is a power-law function of galaxy mass. The result is plotted in Figure \ref{fig:eta}, where we convert each measurement of [Mg/H] to a mass-loading factor. We apply the MC technique to the probability distribution of [Mg/H] directly to obtain the best linear fit to the relation between $\log \eta$ and log mass:
\begin{equation}
    \log\eta = (-0.21\pm0.09)M_{10}-(0.26\pm0.07)
\end{equation}
The mass-loading factor scales with galaxy stellar mass, $\eta \propto M_*^{-0.21\pm0.09}$. A galaxy that is 10 times less massive would have approximately 1.6 times larger outflow rate per SFR.

We note that in fitting the linear relation between $\log \eta$ and log mass, we follow the convention of using fixed values for the yield and the return mass fraction. We exclude 18 galaxies whose best estimated [Mg/H]'s are larger than the adopted yield. Although the yield is found to be quite independent of metallicity, it can vary with different IMFs and the upper mass cutoffs of the IMF \citep{Vincenzo2016}. If we instead measure $\eta$ with an extremely high yield (6 times the solar abundance of Mg), we do not have to exclude those 18 galaxies. The resulting power-law index would be lower ($\eta\propto M_*^{-0.13\pm0.05}$), but is still consistent with the current estimate. The normalization would be $\sim0.5$ dex higher. 

We also experiment on using a different value of return mass fraction. We reduce the return mass fraction to $R=0.23$ to be roughly consistent with the return mass fraction when the \citet{Salpeter1955} IMF is adopted \citep{Vincenzo2016}. The power-law index does not change ($0.21\pm0.09$) but the normalization drops to $-0.11\pm0.07$ dex. In short, choices of the yield and the return mass fraction do not seem to change the resulting power-law index significantly, but they can change the normalization by a factor of a few.

We further explore the massive galaxies in Figure \ref{fig:eta} with very low mass-loading factors. There are a total of 13 galaxies whose measured mass-loading factors lie more than 0.8 dex below the best-fit line. They almost appear to be closed-box systems since, using the adopted yield values, we estimate their mass-loading factors to be less than 5\%. These galaxies are generally massive,  $M_*>10^{10.5}M_\odot$. Their mean [Fe/H] ($0.03\pm0.11$ dex) and age ($3.8\pm2.1$ Gyr) are not significantly different from those of other galaxies also with $M_*>10^{10.5}M_\odot$ (whose mean [Fe/H] and age are $-0.04\pm0.12$ dex and $5.1\pm2.6$ Gyr respectively). However, they are more [Mg/Fe] enhanced (at the mean [Mg/Fe] of $0.42\pm0.11$ dex for those galaxies as compared to the mean of $0.23\pm0.14$ dex for other galaxies). This suggests that they had shorter formation timescales. Moreover, they appear to be near the center of the clusters; most are at less than half the virial radius in projection. It may be interesting to subsequently study in detail the star-formation histories of these galaxies.

\subsection{Comparing the measured mass-loading factors with other measurements and with theoretical predictions}
\label{section:massloading2}
We proposed an alternative way to constrain the mass-loading factor using the metallicities of quiescent galaxies. Our results, using the fiducial yield and the fiducial return mass fraction, agree reasonably well with other works that use chemical evolution models for star-forming galaxies. In Figure \ref{fig:eta}, we plot as a blue dashed line the constraints from \citet{Lu2015}, whose model is very similar to ours (and actually was the model that inspired our work). Their results are upper limits because of the uncertainties in gas fractions, different gas-phase metallicity calibrations, and a conservative choice of yield (5 times solar). Our estimates of $\eta$ are consistent with the upper limit. We also plot an earlier estimate from \citet{Spitoni2010} (in gray shading). Their estimate is based on a chemical evolution model of the gas-phase MZR that allows both infall and outflow.     

Recently, \citet{Belfiore2019} applied chemical chemical evolution models to gas-phase metallicity gradients measured from nearby star-forming galaxies in the MANGA survey \citep{Belfiore2017,Bundy2015}. Their favored model assumes the inside-out growth formalism and that the star formation efficiency is inversely proportional to the orbital timescale. We plot their mass-loading factors as gray hourglass data points in Figure \ref{fig:eta}. The light-gray hourglasses are obtained by adopting the \citet{PettiniPagel2004} O3N2 metallicity calibration, while the dark-gray hour glasses are obtained by adopting the \citet{Maiolino2008} R23 calibration. Our measured mass-loading factors are more consistent with the results using the former metallicity calibration. Their slopes ($-0.58\pm0.20$ and $-0.29\pm0.07$ for the two calibrations respectively) are nominally steeper than our estimation but are consistent with ours within uncertainties.


Our estimates for the mass-loading factors are also consistent with mass loading factors measured in individual galaxies currently driving outflows. Another way to constrain mass-loading factor is to directly observe outflows, via emission lines from the outflow gas or absorption features by the outflow material seen in galaxy or background quasar spectra. From these features, outflow velocities can be determined quite directly. However translating them to mass outflow rate requires a number of assumptions on the ionization structure and the geometry of the outflow. 
The results from applying these techniques to local star-forming galaxies are still limited in terms of sample size. Nonetheless, they suggest a range in mass-loading factor from 0.1 to several tens \citep[e.g.,][]{Bolatto2013, Bouche2012}. In figure \ref{fig:eta}, we plot in navy circles the results based on UV absorption lines from \citet{Chisholm2017}, in which stellar mass measurements were readily available. Those results agree with ours not only in terms of slope but also normalization. In other words, our archaeologically derived mass-loading factors, which are rather of time-average quantities, are consistent with the instantaneous mass-loading factors that are measured in star-forming galaxies.

Given the simplicity of our chemical evolution model, the estimated power-law index of the relation between $\eta$ and $M_*$ agrees surprisingly well with both analytic and hydrodynamical simulations. \citet{Hayward2017} presented an analytic model for galactic outflows driven by stellar feedback. In the model, the SFR is self-regulated. It is set by an equilibrium between the momentum injection rate from stellar feedback and the dissipation rate by turbulence and occasional outflows. The outflow occurs when the momentum accumulated within a coherence time is large enough given the gas density of the ISM patch. Their model predicts that the mass-loading factor scales with gas fraction and stellar mass. For the stellar mass component, their predicted scaling relation is $\eta \propto M_*^{-0.23}$, which is consistent with our observation.

However, the observed exponent of the scaling relation is smaller than predicted in an earlier semi-analytical model by \citet{Lagos2013}. In this model, the basic assumptions are similar to those of \citeauthor{Hayward2017}: the ISM disk is supported by turbulent pressure, and stellar feedback drives gas outflow. However, the timescale of the \citeauthor{Lagos2013} simulation is set by the lifetime of giant molecular clouds, which can be larger than the crossing time adopted by \citeauthor{Hayward2017}, and their SFR is set by an empirical relation. They also found that the mass-loading factor scales with gas density and galaxy mass with an exponent of $\sim-1.1\pm0.5$ (based on estimation from their Figure 14), which is larger than what we observed.

Hydrodynamical simulations predict scaling relations of the mass-loading factor that are consistent in slope but with somewhat larger normalization than the observational result. The dotted line in figure \ref{fig:eta} shows the scaling relation presented by \citet{Muratov2015} based on the Feedback in Realistic Environments (FIRE) simulations. The slope is consistent with our observation but it is $\sim3-7$ times larger in normalization than our best-fit line. 

It is normally difficult for simulations to define mass-loading factors that match observations. \citet{Muratov2015} calculated the mass-loading factors from outflow flux (all gas with outward radial velocity) at $0.25R_{vir}$. This does not capture the effect of recycling from galactic fountains, whereas our measurement does. Thus, their results are likely to be larger than our measurements by construction. Based on the IllustrisTNG results from \citet{Pillepich2018}, in which their definition of the mass-loading factor was explicitly remarked as not applicable to observations and was calculated from wind energy at injection, their ``mass-loading factor'' scales with halo mass $M_{200}$ with the power law index of $\sim-0.8$ to $-0.6$. If we convert halo mass to stellar mass using the stellar-to-halo mass relation $M_*\propto M_{200}^{\sim2.27}$, found in the same suite of simulations \citep{Niemiec2018} for galaxies with $M_*<10^{11.2}M_\odot$, the power-law index with stellar mass is approximately $-0.35$ to $-0.26$, which is still roughly consistent with what we measured.

\subsection{Remarks on the MZR evolution with redshift}
\label{sec:discuss_evol}
\begin{figure*}
    \centering
    \includegraphics[width=0.497\textwidth]{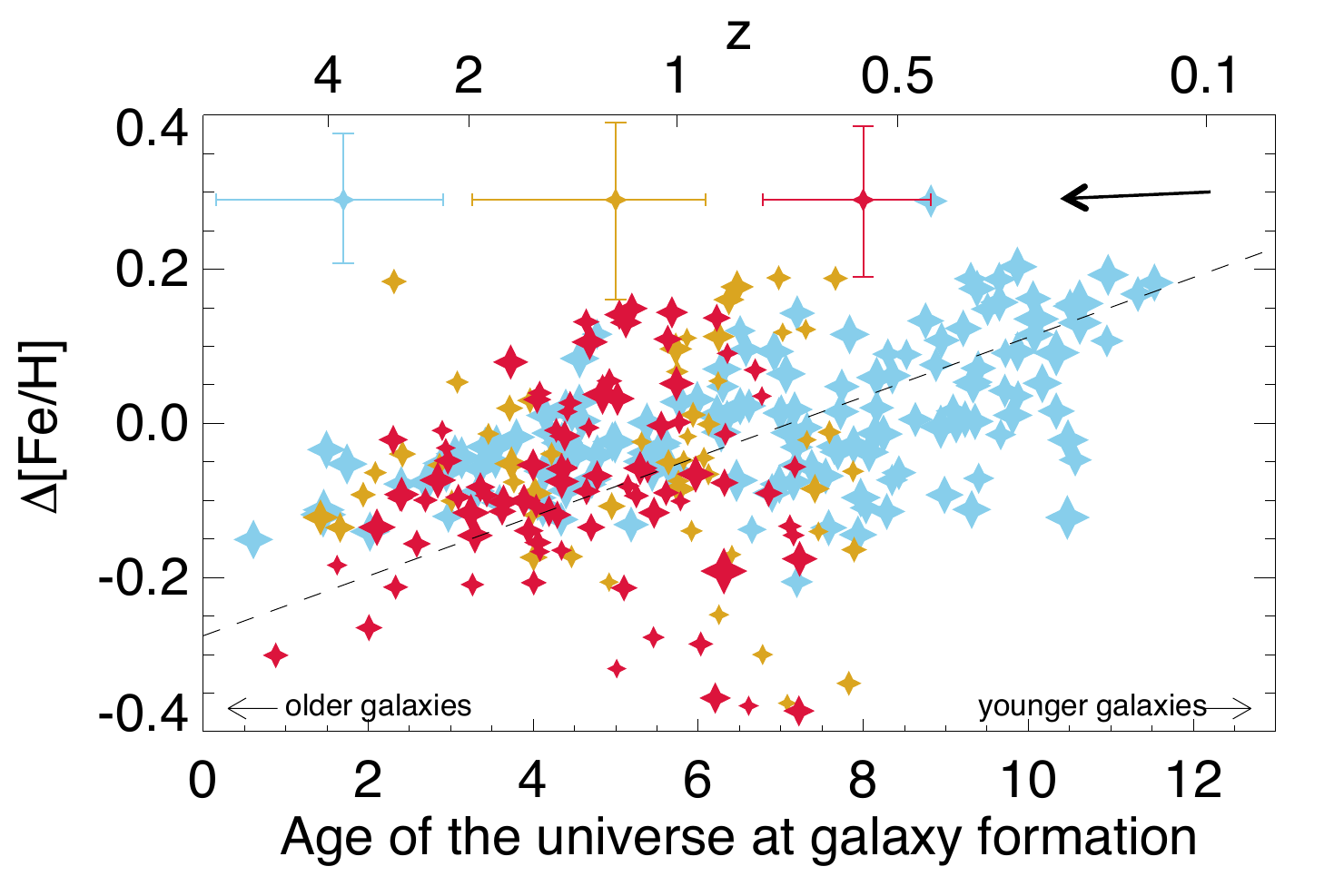}
    \includegraphics[width=0.497\textwidth]{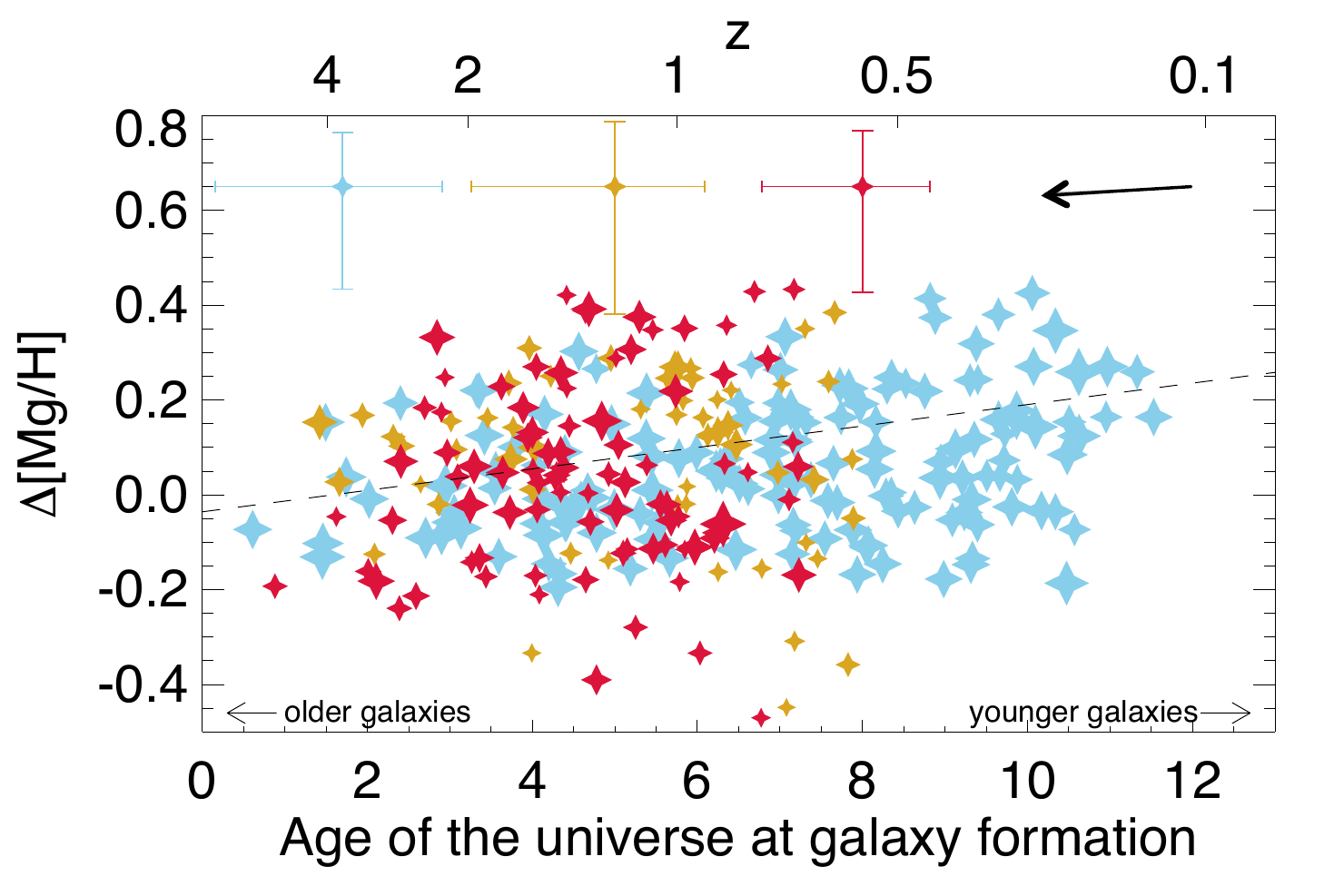}
    \caption{The dependence of [Fe/H] and [Mg/H] on formation redshift when the dependence on mass is removed (see Section \ref{sec:discuss_evol}). The median uncertainties are shown in the top of each panel. The color scheme follows previous plots. The slope on the left is statistically significant, but the slope on the right is not. The arrows at the top right corners show the slopes in the age-metallicity planes that are induced by the age-metallicity degeneracy.}
    \label{fig:formationtime}
\end{figure*}
In Paper I, we argued that the evolution of metallicity with formation redshift in quiescent galaxies is more fundamental than the evolution with observed redshift. Galaxies of the same mass that formed at the same redshift should have similar metallicities regardless of when we observe them, assuming that most quiescent galaxies evolve passively.\footnote{If the sample size were large enough, we could also use this technique to trace whether quiescent galaxies evolve passively in metal abundance. This could be done by comparing MZRs of galaxies that formed at the same formation redshift but are observed at different redshifts.} As in Paper I, we trace each galaxy back to its formation time by adding the age of the universe when it is observed to the age of the galaxy. We plot the deviation in metallicities from the $z\sim0$ MZR (Equation \ref{eq:explicit_mzr}) as a function of formation time in Figure \ref{fig:formationtime}. The plots essentially show the dependence of metallicities on formation redshift (age), while the dependence on mass is removed.

The deviation in the [Fe/H] abundances is consistent with what we found in Paper I. The [Fe/H] evolution with formation time is $0.04\pm0.01$ dex per formation Gyr. However, we do not find a significant dependency of [Mg/H] on formation time. As shown in the right of Figure \ref{fig:formationtime}, the best-fit slope for the evolution is $0.02\pm0.01$ dex per formation Gyr, a significance less than $2\sigma$ at face value. 

We further estimate the impact of the age-metallicity degeneracy on the evolution of [Fe/H] and [Mg/H] with formation time. In a similar manner as that in Paper I, we create a set of mock observations by assuming that each galaxy of the observed mass and age has the intrinsic [Fe/H] and [Mg/H] values that follow the $z\sim0$ MZR, i.e., hypothetically there is no evolution with redshift. We create SSP spectra with these ages and metallicities, add Gaussian noise with the same flux uncertainty array as the observed spectra, and fix the velocity dispersion to 250 km/s (to keep the computation time reasonable). We then calculate a $\chi^2$ grid for each spectrum by comparing to the noiseless grid used to calculate systematic uncertainties in Section \ref{sec:modelfitting}. The ``observed'' age, [Fe/H] and [Mg/Fe] of each mock galaxy is then selected according to the probability from the $\chi^2$ grid. These mock galaxies do not have any intrinsic deviation from the $z\sim0$ MZR. Therefore, if there is any evolution with formation time, it is caused by the age-metallicity degeneracy. 

Based on these mock observations, we find that the age-metallicity degeneracy cannot explain the dependence of $\Delta$[Fe/H] on formation time and even lowers the significance for the dependence of $\Delta$[Mg/H] on formation time. Similar to what we found in Paper I, the age-metallicity degeneracy induces a small positive slope of [Fe/H] at $0.004\pm0.002$ dex per formation Gyr. For [Mg/H], it induces $0.010\pm0.004$ dex per formation Gyr. These slopes are illustrated at the top-right corner of each panel in Figure \ref{fig:formationtime}. This confirms that the evolution of [Fe/H] with formation time is significant at $>3\sigma$ but the evolution of [Mg/H] is less than $2\sigma$ significant. An evolution of [Mg/H] with formation time would imply that fundamental properties of galaxies such as yield or mass-loading factors depend on time. Since we did not detect any significant evolution of [Mg/H] with both observed and formation redshift, these fundamental properties are not required to depend on redshift.


The absence of the evolution of [Mg/H] with redshift together with the existence of the evolution of [Fe/H] with redshift suggest that the latter is caused by the delayed time of the Type Ia supernovae. The main difference between the two elements is that Mg is approximately instantaneously recycled while the Fe can have a delay time up to several Gyrs \citep[e.g.,][]{Maoz2012}. Thus, the evolution of [Fe/H] with redshift in quiescent galaxies that we found is not surprising and is perhaps expected due to selection effects. We restricted our sample to be passive. The galaxies at higher redshift are required to finish forming star early, while the galaxies at lower redshift could have quenched at a later time. Quiescent galaxies at higher redshift had less time to become enriched by Fe from Type Ia supernovae. 

\begin{figure}
    \centering
    \includegraphics[width=0.497\textwidth]{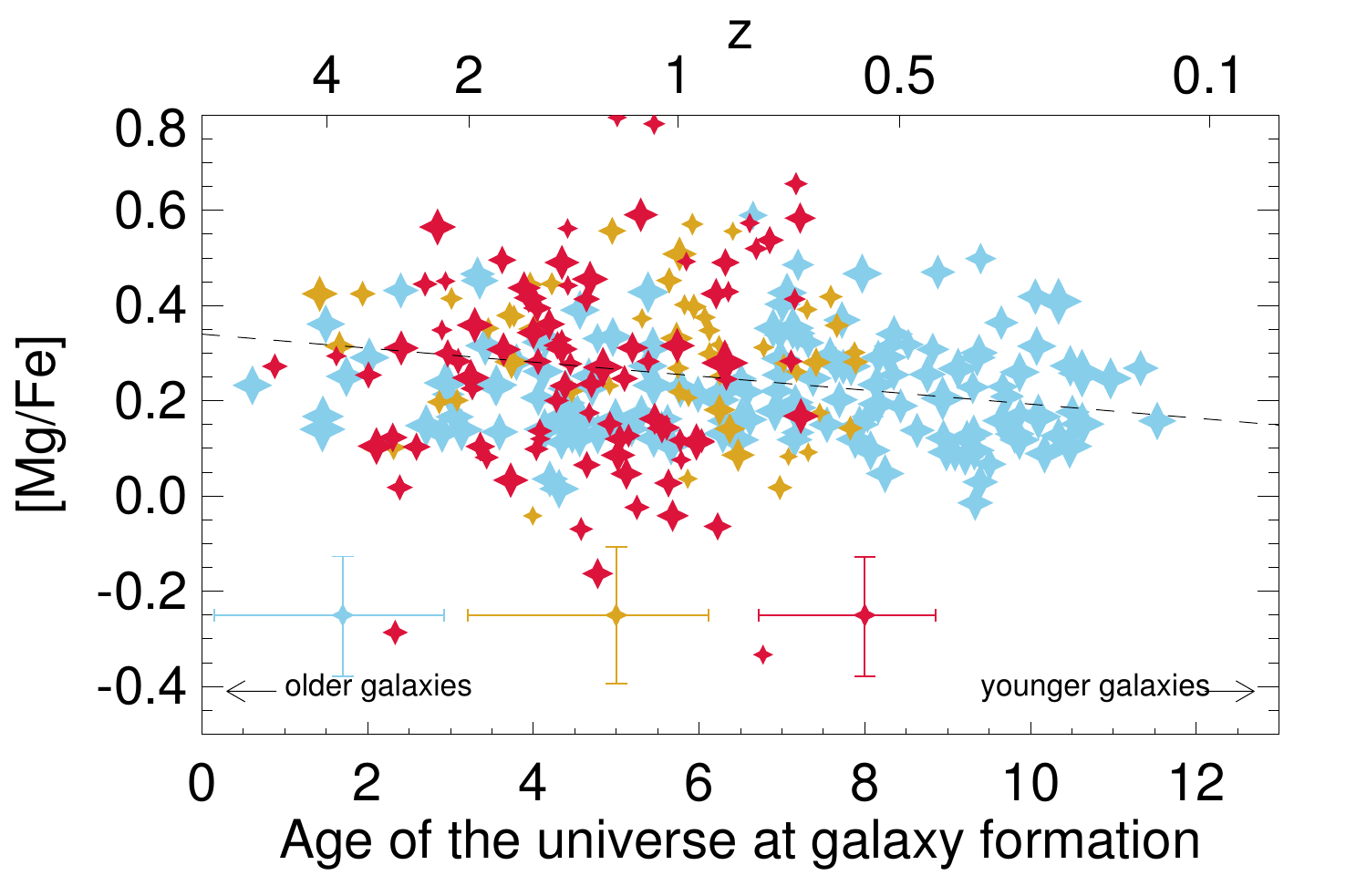}
    \caption{The dependence of [Mg/Fe] on formation redshift. The median uncertainties are shown in the bottom. The color scheme follows previous plots (indicating different observed redshifts). A slight negative slope suggests that quiescent galaxies that form earlier on average had shorter star formation duration.}
    \label{fig:mgfe_tform}
\end{figure}

We plot [Mg/Fe] as a function of formation time in Figure \ref{fig:mgfe_tform}. This is to further investigate the conjecture that the evolution of [Fe/H] with redshift is due to shorter star formation histories of quiescent galaxies at an earlier time. Since the ratio between alpha elements and Fe is an indication of star formation duration \citep{Thomas2005}, we expect that, on average, galaxies that form earlier have higher [Mg/Fe]. The slope in Figure \ref{fig:mgfe_tform} is $-0.015 \pm 0.006$ dex per formation Gyr, suggesting that galaxies that form earlier indeed have higher [Mg/Fe] than galaxies that form later. The slope is also as expected,roughly equal to the difference in the slope in the $\Delta$ [Mg/H] with formation time and the slope in the $\Delta$ [Fe/H] with formation time in Figure \ref{fig:formationtime}. 

The evolution of [Fe/H] with redshift we found in this work does not necessarily contradict the non-detection of evolution of MZR with redshift in \citet{Gallazzi2014,Estrada-Carpenter2019,Saracco2019}. This is mainly due to large measurement uncertainties. We first note that most of these works measure total metallicity [Z/H] without $\alpha$ enhancement. However, most of the spectra either do not cover the Mg b lines, or the absorption indices were chosen to be less sensitive to $\alpha$ enhancement. Therefore these [Z/H] should be close to our [Fe/H] measurements. Regardless, \citet{Gallazzi2014}, \citet{Estrada-Carpenter2019} and \citet{Saracco2019} found that the [Z/H]--mass relations in quiescent galaxies (both in the field and in clusters) at $z\sim0.7-1.6$ are not different from that of local galaxies. These studies are quite limited in terms of sample size, especially for the lower mass galaxies, and either small wavelength coverage or coarse spectral resolution. These aspects result in large measurement uncertainties that can disguise the expected evolution.

For example, based on the mean age of $2.0\pm0.8$ Gyr of 7 quiescent galaxies at $z\sim1.22$ in \citeauthor{Saracco2019}'s (\citeyear{Saracco2019}) sample and the amount of evolution with formation time found in this work, we expect them to have an average $0.15\pm0.06$ lower metallicity than the local galaxies. This number is smaller than their mean uncertainty in [Z/H] measurements (0.22 dex), standard deviation within the sample (0.18 dex), and uncertainty in the normalization of the best-fit linear relation with stellar mass (0.15 dex).

\subsection{Limitations of our approach}
The formalism for the stellar metellicities of quiescent galaxies adopted here is simple and subject to certain limitations. Aside from those listed in the explicit assumptions regarding the metallicities of inflow and outflow, which are discussed in Appendix \ref{appendix:enrichedoutflow}, there are additional limitations involved in simplifying Equation \ref{equation:chem2} to Equation \ref{equation:quiescentz}. 

First is the rapid truncation of star formation by non-star forming activities when the galaxy is still gas rich, i.e., $r_g$ may not be as negligible as assumed. This includes active galactic nuclei (AGN) feedback and ram-pressure stripping. It is still unclear whether AGNs are able to eject significant mass fractions out of the galaxies. There is evidence that AGNs can drive large outflow in individual cases \citep[e.g.][]{Maiolino2012,Nyland2013}, or in more extreme objects such as ULIRGs and QSOs \citep[e.g.][]{Cicone2014,Rupke2017}. Nevertheless, many statistical studies found that AGN-driven winds are usually not strong enough to escape the galaxies \citep[e.g.][]{Sarzi2016,Concas2017,RobertsBorsani2019}. We defer the discussion on ram-pressure stripping to the next section on galaxy environment. 

Moreover, the formalism adopted here does not consider mergers. The most concerning case is a dry merger, as it increases the stellar mass but does not increase the stellar metallicity. There is evidence that the most massive galaxies have experienced multiple minor mergers in their outskirts, which is responsible for their size growth \citep[e.g.][]{VanDokkum2010,Greene2012,Greene2013}. The effect of this is beyond our study. Nonetheless, we can consider a simpler case of a major merger of two quiescent galaxies with equal masses and metallicities that lie on the MZR\@. In this case, the descendant galaxy will double in stellar mass (an increase of 0.3 dex), while the metalllicity remains the same. However, we estimate that the effect in this case is quite small. Because the MZR of quiescent galaxies has a gentle slope ($\lesssim0.2$ dex per log mass in our finding), the descendant galaxy of a major merger will have a metallicity that is just $\lesssim0.06$ dex below the MZR relation.    


\subsection{Effect of galaxy environment}
As already discussed in Paper I, we do not expect our finding to be affected by the environment. In short, a cluster environment mainly affects the quenched fraction of galaxies in the sense that cluster galaxies tend to be more quenched than field galaxies. However, in terms of chemical composition, the effect is either small \citep[$<0.06$ dex, e.g.,][]{Cooper2008,Jorgensen2018}, or not significant \citep[e.g.,][]{Harrison2011,Kacprzak2015,Fitzpatrick2015}.

These findings may be relevant to the role of ram-pressure stripping in galaxy clusters. Based on the SDSS group/cluster catalogues and a cosmological simulation, \citet{Wetzel2013} suggest that most satellite galaxies can retain their molecular gas in the disks after falling into clusters. These galaxies can continue their star formation histories in a similar manner (within 10\%) to those of central galaxies. The authors suggest that this is consistent with the observations of molecular gas in the satellites of the Virgo cluster \citep{Kenney1986,Young2011}. Moreover, based on a different cosmological simulation, \citet{Bahe2015} find that the role of ram-pressure stripping is most relevant to the removal of the atomic gas in the outer halo just before the star formation stops.

\section{Conclusion} \label{sec:conclusion}

We measured ages and metallicities ([Fe/H] and [Mg/H]) of individual quiescent galaxies in two galaxy clusters at $z\sim0.39$ and $z\sim0.54$ using full-spectrum fitting. The sample consists of 62 galaxies at $z=0.39$ and 92 galaxies at $z=0.54$ in the galaxy clusters Cl0024 and MS0451. We also used a subsample of 155 SDSS quiescent galaxies as reference $z\sim0$ galaxies. We used the SPS models and response functions from \citet{Conroy2009} and \citet{Conroy2018} to fit the observed spectra. By expanding the sample size from \citet{Leethochawalit2018} and measuring Mg abundances in addition to Fe, we were able archaeologically to derive the dependence of mass-loading factor on mass based on the mass--metallicity relation of quiescent galaxies. The summary of the findings this paper are the following.

\begin{itemize}
    \item We confirmed the finding in Paper I that the stellar mass--[Fe/H] relation evolves with redshift. At a fixed stellar mass, quiescent galaxies at higher redshift have lower [Fe/H] than quiescent galaxies at lower redshift. The slope of the relation is also likely steeper at higher redshift.
    \item We found no evolution in the mass--[Mg/H] relation with redshift. This suggests that the evolution observed in the [Fe/H] abundance is due to the delay time of Type Ia supernovae.  
    \item We constrained the mass-loading factor using an analytic chemical evolution model for quiescent galaxies, assuming that Mg is an indicator of instantaneously recycled elements. We found that the mass-loading factor is a power-law function of galaxy stellar mass, $\eta \propto M_*^{-0.21 \pm 0.09}$, over the observed mass range $\sim10^{9.5}$ to $10^{11.5} M_\odot$.  
    \item Our constraint on the mass-loading factor is consistent with an analytic prediction in which outflow is caused by star-formation feedback in a turbulent disk \citep{Hayward2017}. It is also consistent with the results from direct measurements of outflow using UV absorption lines.  
\end{itemize}

There are still many open questions that can be explored beyond this paper. These include confirming the stellar mass--metallicity relationship (in both [Fe/H] and alpha elements) with a large sample of field galaxies, especially down to low masses ($\sim10^{9.5}~M_\odot$). A large sample of abundance measurements in field quiescent galaxies at different redshifts would allow us not only to even more firmly establish the evolution of MZRs, but also, as suggested in Section \ref{sec:discuss_evol}, to determine the degree of passive evolution of quiescent galaxies in terms of their chemical composition. This can be done by comparing the MZRs of galaxies that formed at the same redshift but are observed at different redshifts. Beyond this, precise and consistent measurements of stellar metallicities in star-forming galaxies (especially as a function of gas fraction), will help bridge and strengthen the findings in our work with those by \citet{Gallazzi2005} and \citet{Lu2015}.

\acknowledgments
We thank the referee for comments and suggestions that significantly improved the paper. The authors acknowledge Charlie Conroy, Philip Hopkins, and Gwen Rudie for useful feedback and the use of their models and fitting codes.

\facilities{Keck:II (DEIMOS), Sloan, GALEX, HST (WFPC2), Subaru (Suprime-Cam), CFHT (CFH12k), Hale (WIRC)}

\software{KCORRECT \citep{Blanton2007}, FSPS, \citep{Conroy2009}, alf \citep{Conroy2018}, MPFIT \citep{Markwardt2012}}

\appendix
\section{Choosing the best combination of response functions}
\label{sec:testcomb}
Although the FSPS code is able to measure many elemental abundances simultaneously, it requires more computation time to measure more elements.  Every additional element requires one additional spectral response function.  To achieve reliable measurements of [Mg/Fe] within a reasonable computation time, we tested several combinations of response functions to be included in the fitting models. We performed these tests on two sets of data. First, we used our fitting code to fit the high-S/N, stacked spectra of \citet{Choi2014}. We then compared our results to the \citeauthor{Choi2014}'s measurements. Second, we used the full-spectrum fitting code \texttt{alf} \citep{Conroy2018} to re-measure abundances for a subset of our sample.  We compared the \texttt{alf} measurements to the results described according to the method described in Section \ref{sec:modelfitting}.

\subsection{Comparison with Choi14}
\label{section:comparechoi}
 In this section, we test our measurements of age and metal abundance on 10 high-S/N stacked spectra from \citet{Choi2014}. The spectra were stacked from individual spectra in the AGN and Galaxy Evolution Survey \citep[AGES,][]{Kochanek2012} based on their redshifts, $z=0.3$ to 0.7, and masses, $M_*=10^{10.2}$ to $10^{11.3} M_\odot$. The spectral resolution is 6 \AA\ with the wavelength range of 4000 to 5500 \angstrom\@. 
 
 We consider 5 spectral models with different sets of response functions in addition to the four baseline parameters: age, [Z,H], velocity dispersion, and redshift. The combinations are
 
 \begin{enumerate}
     \item Mg only,
     \item alpha elements (Mg, Si, Ca, Ti, and O) all fixed to the same value,
     \item Mg and O,
     \item Mg and N,
     \item ``full'' combination having the same list of 9 elements as \citet{Choi2014}: Mg, N, Fe, O, C, N, Si, Ca and Ti.
\end{enumerate}
 
\begin{figure*}
\gridline{\fig{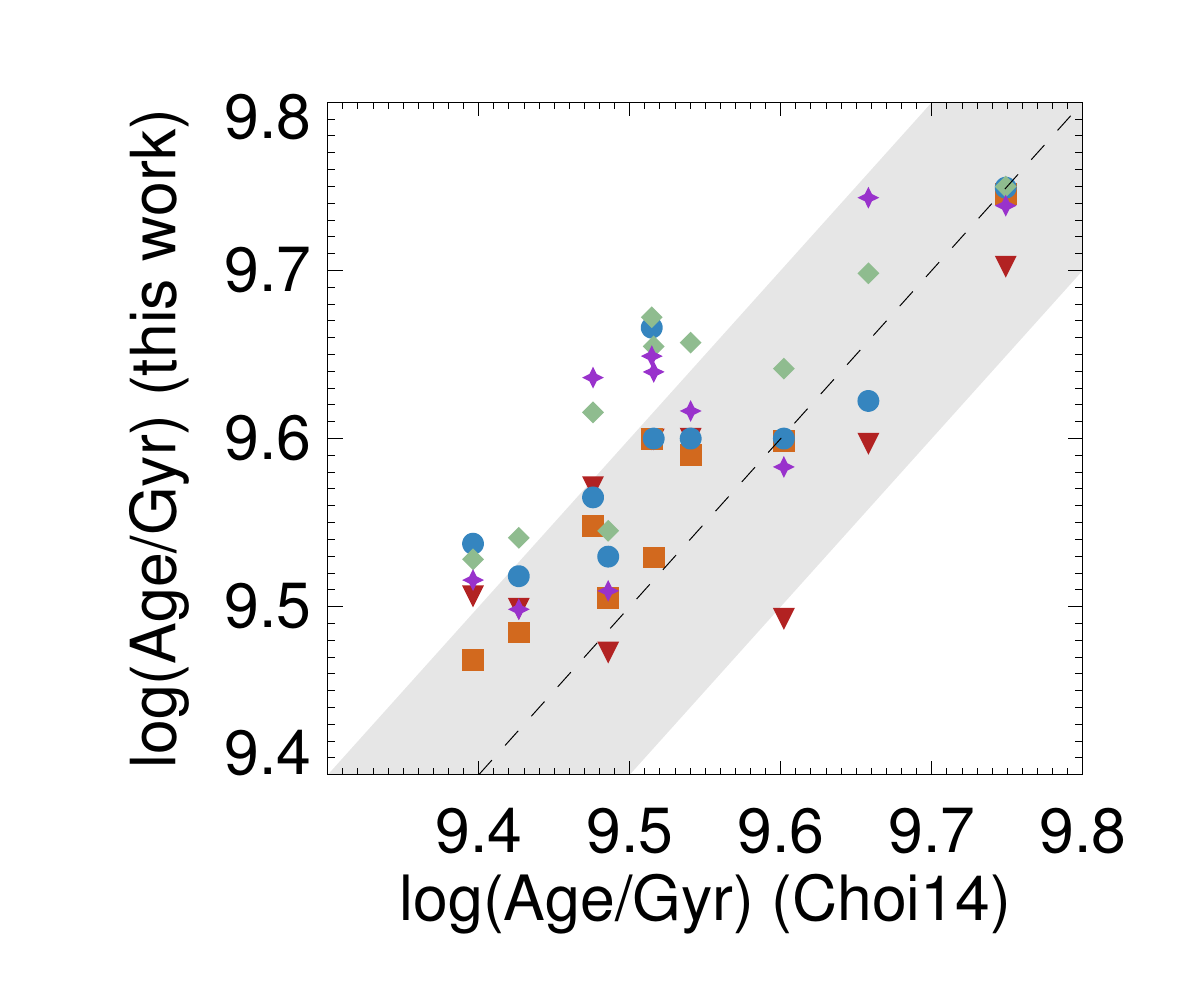}{0.33\textwidth}{(a)}
          \fig{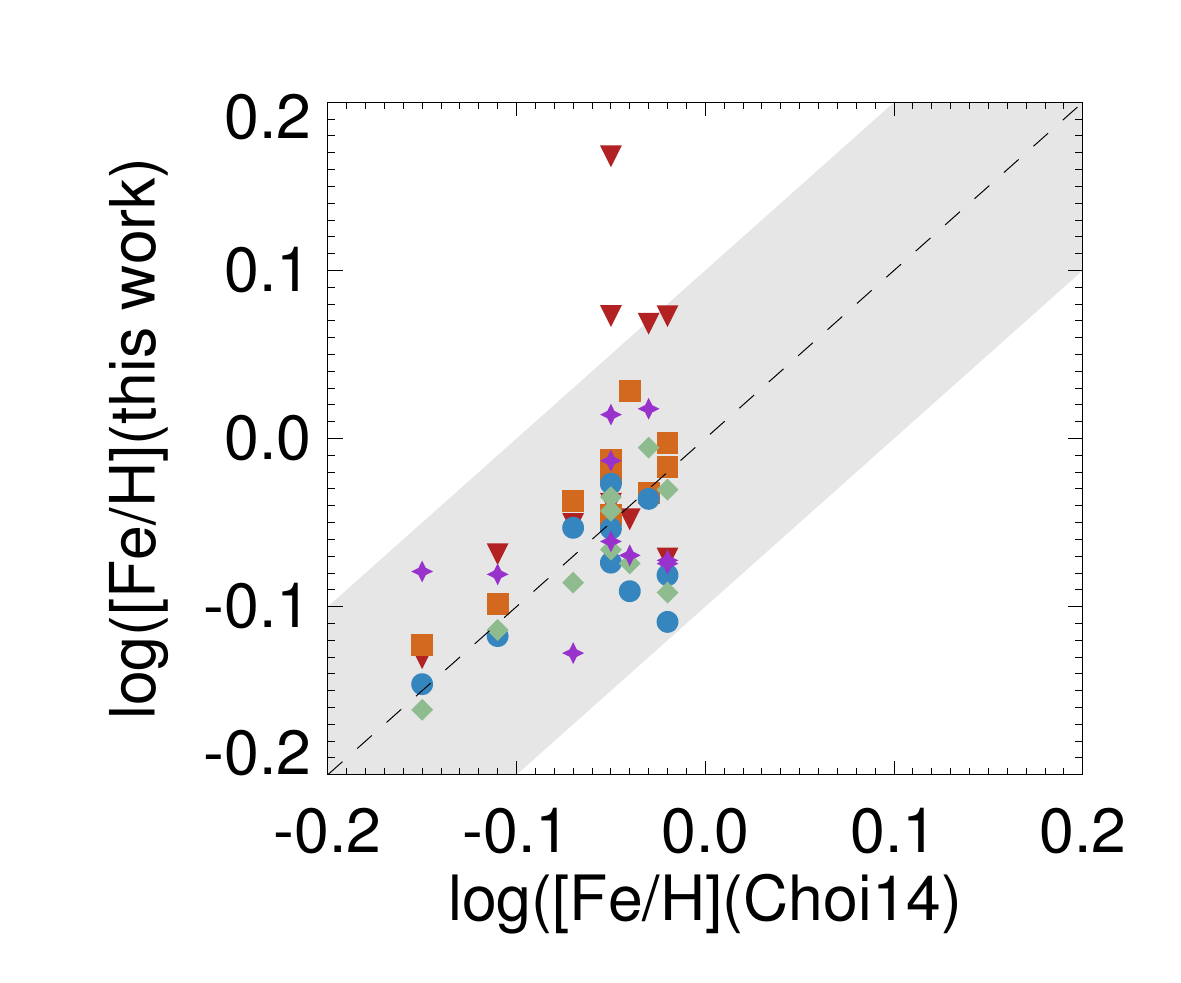}{0.33\textwidth}{(b)}
          \fig{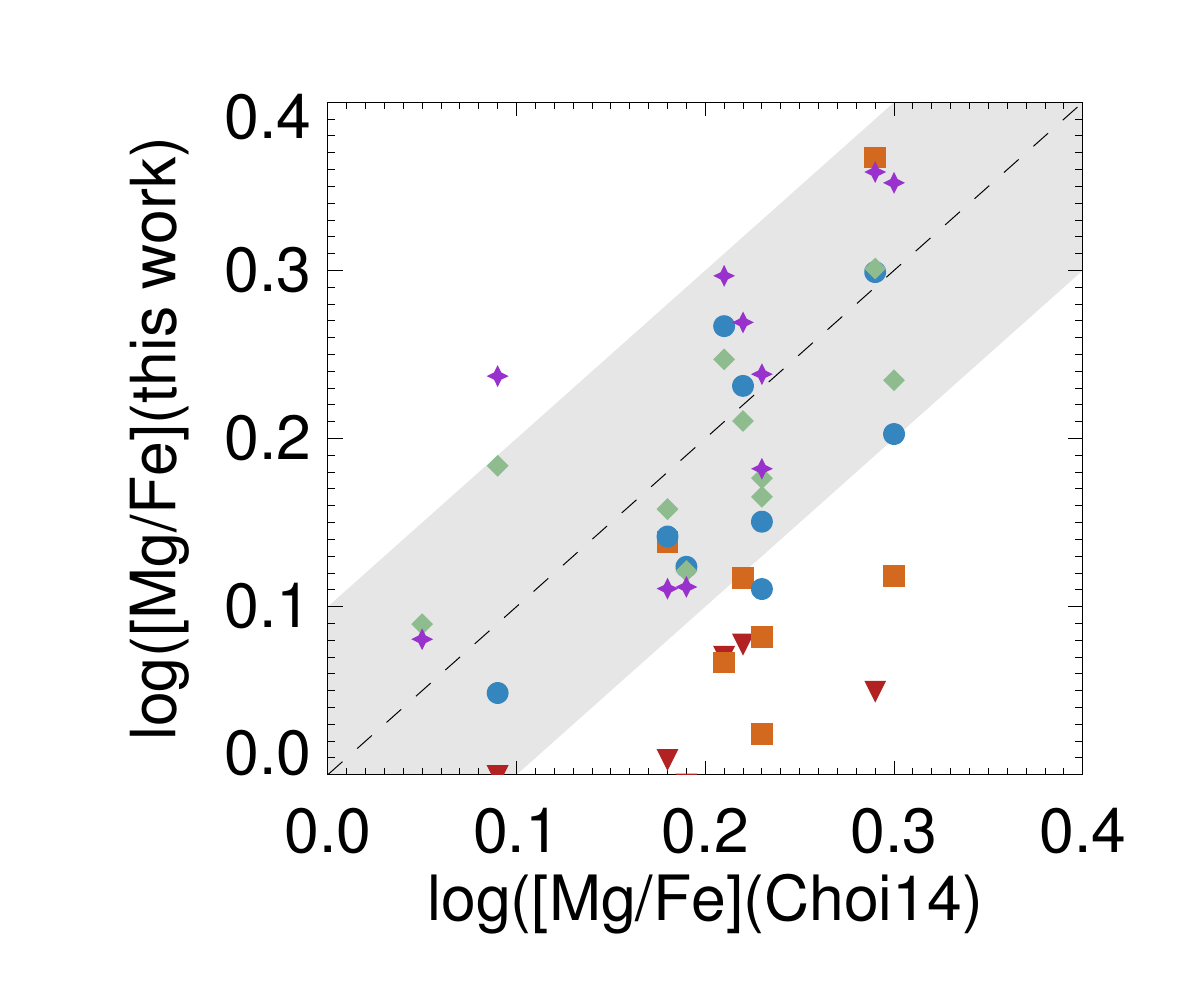}{0.33\textwidth}{(c)}
          } 
\caption{Comparison with the measurements of \citet{Choi2014} using different combinations of metal elements in the response functions. The combinations are (1) [Mg/Fe] only (orange squares), (2) alpha elements [Mg,Si,Ca,Ti,O/Fe] all fixed to the same value (red downward triangles), (3) [Mg/Fe] and [O/Fe] (blue circles), (4) [Mg/Fe] and [N/Fe] (green diamonds), and (5) Mg,N,Fe,O,C,N,Si,Ca and Ti/Fe (purple stars). The latter three combinations show the best agreement with \citet{Choi2014}.     \label{fig:choicompare}}
\end{figure*}

We compare the results from each model with the measurements from \citet{Choi2014} (see Figure \ref{fig:choicompare}). We find that three of five the models agree within 0.1 dex  with the measurements from \citeauthor{Choi2014} These models are those that include the response functions of (3) Mg and O, (4) Mg and N, and (5) the ``full'' combination. 

In terms of the age measurements, all models perform equally well. The agreement holds when the age is older than $\sim 3.5$ Gyr.  For younger populations, the age measurements obtained here are $\sim0.05$ dex older than the ages reported by \citeauthor{Choi2014} This trend of the differences in the age measurements is the same as what we found in Paper I when we fit for only ages and [Z/H]. We attributed the difference to the lack of age-sensitive higher-order Balmer lines ($<$4000\angstrom) in the spectra and possibly the differences in the models used (see Appendix of Paper I).

However, we find agreement in [Fe/H] and [Mg/Fe] with \citeauthor{Choi2014} only in the three combinations of response functions mentioned above. When the enhancements of all alpha elements are fixed to the same value or when only the [Mg/Fe] response function is included, the [Fe/H] is generally slightly over-predicted while the [Mg/H] is under-predicted (red triangles or orange squares in Figure \ref{fig:choicompare}). The former case (all alpha elements fixed to the same value) is not surprising because alpha elements generally do not have the same values, and they do not always track each other.  \citet{Conroy2014} found that Mg tracks O closely, but Mg does not track heavier $\alpha$ elements such as Si, Ca, and Ti. 
The latter case (Mg response function only) is more interesting. We cannot measure [Mg/Fe] or [Fe/H] well when we only include the response function of Mg. This is probably because the Mg enhancement not only strengthens the absorption lines in the Mg~b $\lambda$5170 region but also weakens the absorption lines at $\sim4000$--4400 \angstrom\@. The wavelength range also responds to the enhancement of Fe, O, and N (See Figures 7 and 8 of \citealt{Conroy2018}). For this reason, the agreements improve to within 0.1 dex when we include the response functions of either O or N\@.   

In summary, we found that our fitting code yields results that agree reasonably well (within 0.1 dex) with the literature when the we use the following models: Mg and N, Mg and O, or the ``full'' combination. However, the ``full'' combination model, which includes 13 parameters, is costly in terms of the computational time.  Therefore, it is difficult to apply to a large sample. Thus, we restrict further consideration to the two simpler models that include the response functions of either Mg and N or Mg and O\@.

\subsection{Comparison to the \texttt{alf} full-spectrum fitting code}
In order to choose the better of the remaining two models, we compare the results from our fitting code to the results from the publicly available ``absorption line fitting code'' \citep[\texttt{alf},][]{Conroy2018}. \texttt{alf} is a fitting algorithm that uses the Markov Chain Monte Carlo method of parameter estimation. It constructs empirical SSP spectra from the MIST isochrones \citep{Choi2016}, the MILES and Extended IRTF spectra libraries, and theoretical response functions of individual elements based on the Kurucz suite of routines. Our code uses a different set of empirical SSP spectra \citep{Conroy2009} but uses the same response functions. All of the \texttt{alf} results presented here were modeled with the Kroupa IMF and the ``simple'' mode of fitting, which fits for 13 parameters including stellar age, [Z/H], and abundances of Fe, C, N, O, Na, Mg, Si, Ca, and Ti. 

\begin{figure*}
\gridline{\fig{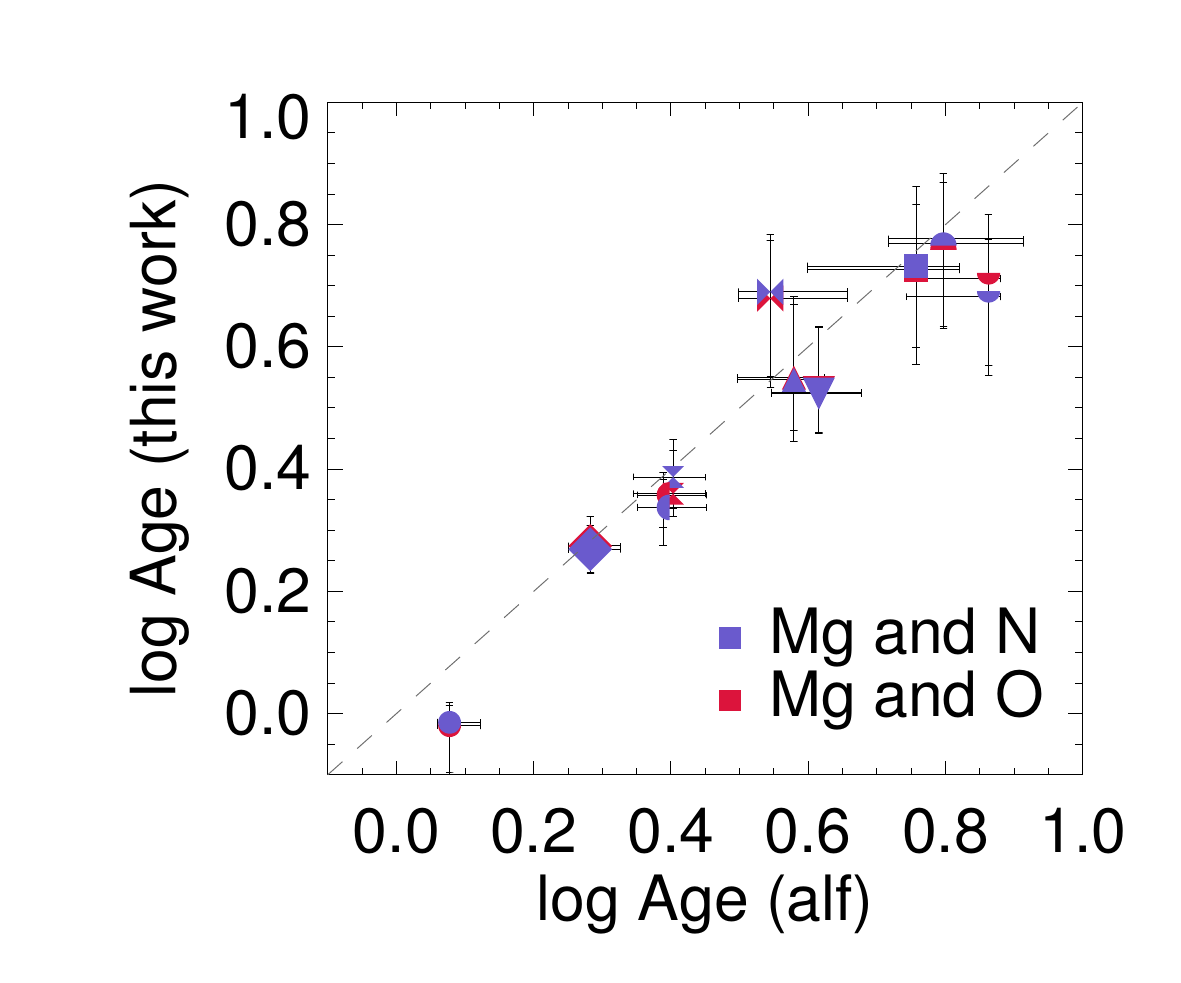}{0.33\textwidth}{(a)}
          \fig{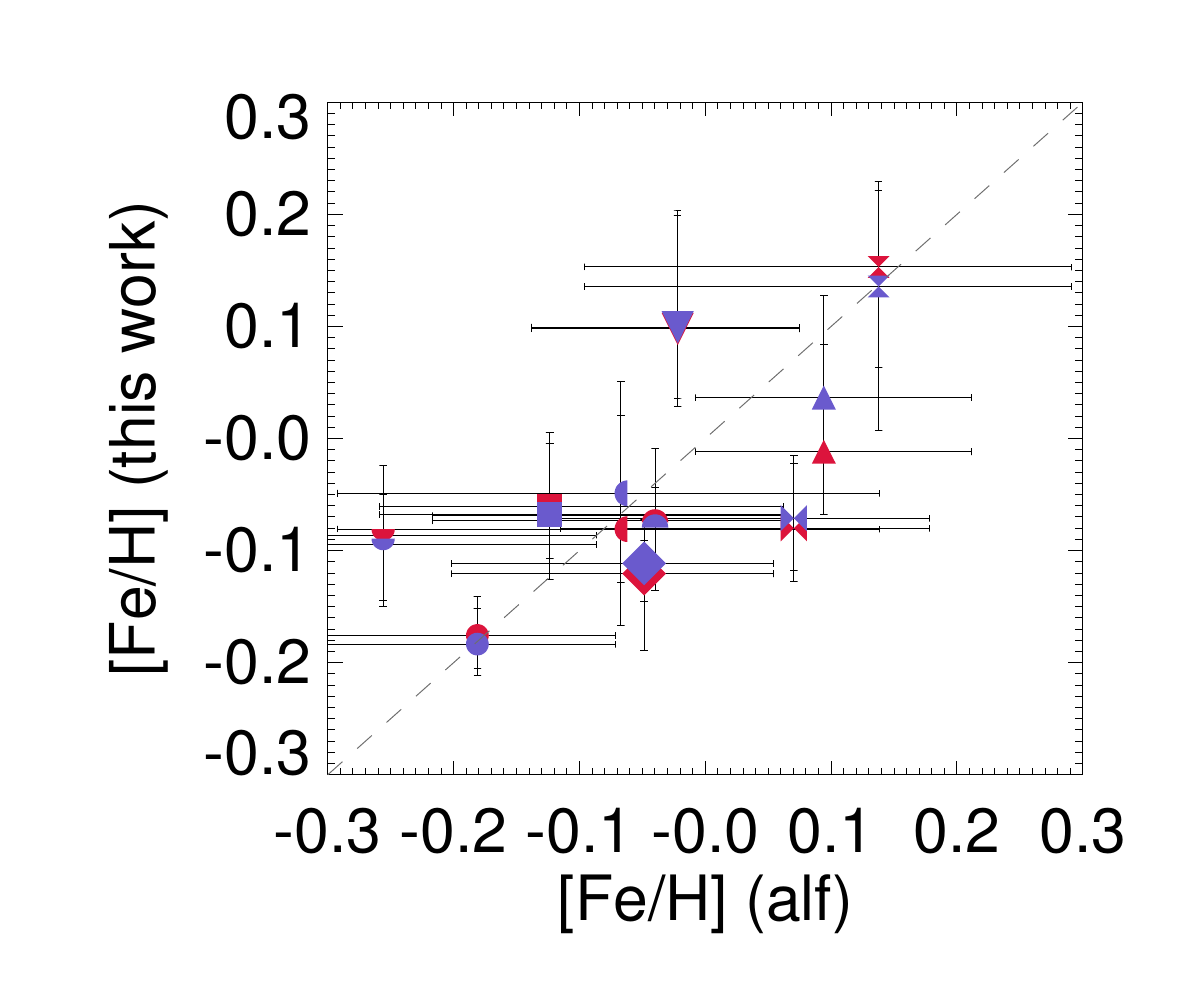}{0.33\textwidth}{(b)}
          \fig{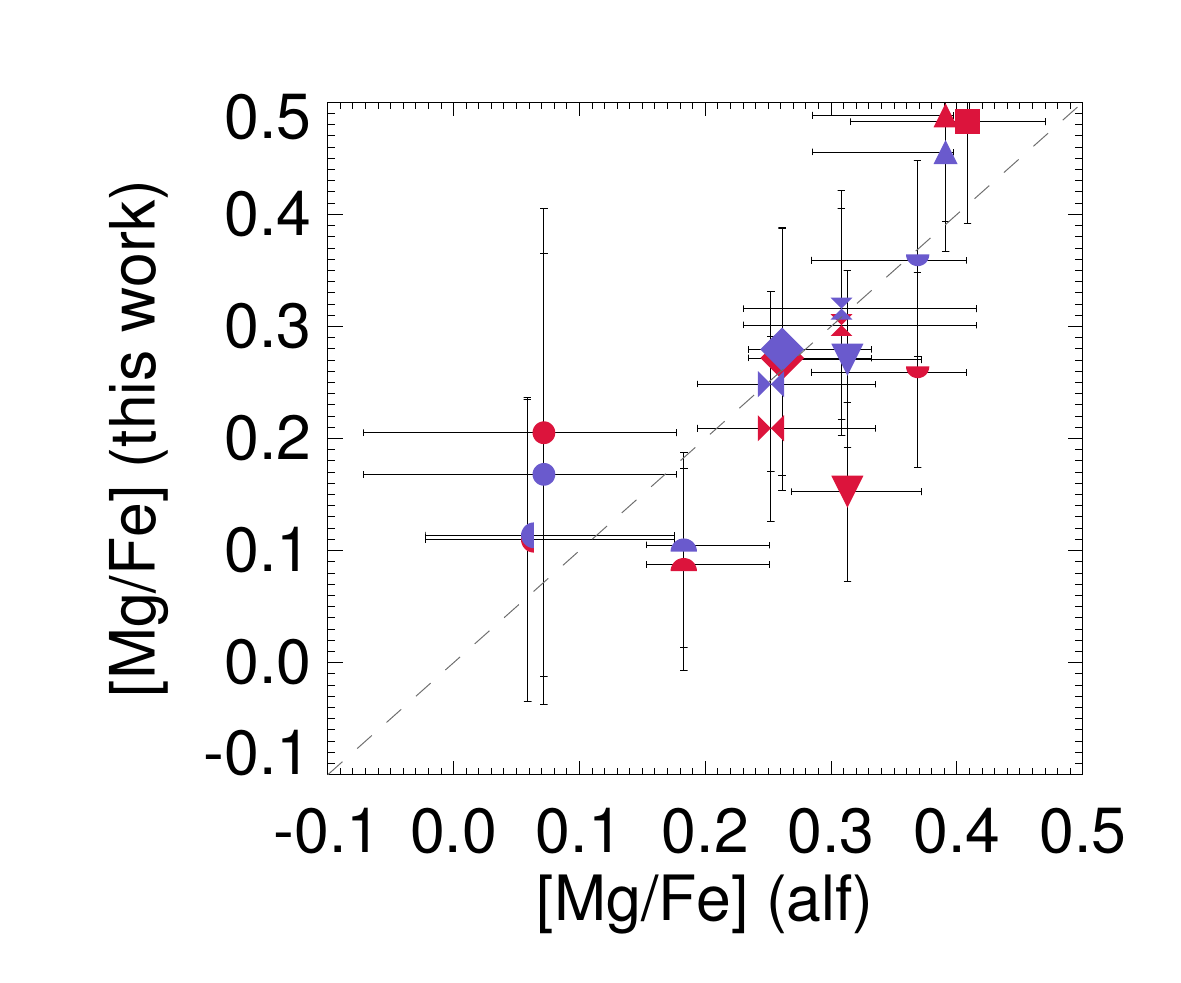}{0.33\textwidth}{(c)}
          } 
\caption{Comparison of  parameters of 10 galaxies measured with \texttt{alf} \citep{Conroy2018}, which fits for the abundances of nine individual elements, and measure with our code using the combination of response functions of (1) [Mg/Fe] and [N/Fe] and (2) [Mg/Fe] and [O/Fe]. We chose combination (1) in our final fitting code to apply to the full sample.   \label{fig:alfcompare}}
\end{figure*}

Because the \texttt{alf} computational time for a single spectrum is long ($\gtrsim 100$ cpu hours), we selected the 10 highest S/N spectra from our sample in MS0451 ($z\sim0.54$). Figure \ref{fig:alfcompare} shows the comparison between the two methods. Our measurements are remarkably consistent with the results from \texttt{alf}, given that our code is simpler. However, We found that in some of the cases, the model that includes the response functions of Mg and N yields better agreement, especially in terms of [Mg/Fe], than the model that includes the response functions of Mg and O\@. Based on the results in this section, we chose the model that includes the response functions of Mg and N to apply to our data.

\subsection{Comparison between the fitting method in Paper I and in this paper}
\label{appendix:paper1compare}
Here we compare the [Fe/H] and age measurements derived from the method in Paper I to those derived in this paper. In Paper I, we masked out the Mg~b absorption region in the spectra and measured [Fe/H] and age with without $\alpha$ enhancement. In this paper, we fit the full spectra with models that treat N and Mg abundances as free parameters. We use both methods to fit the higher-redshift spectra in this paper. 

\begin{figure}
    \centering
    \includegraphics[width=0.49\textwidth]{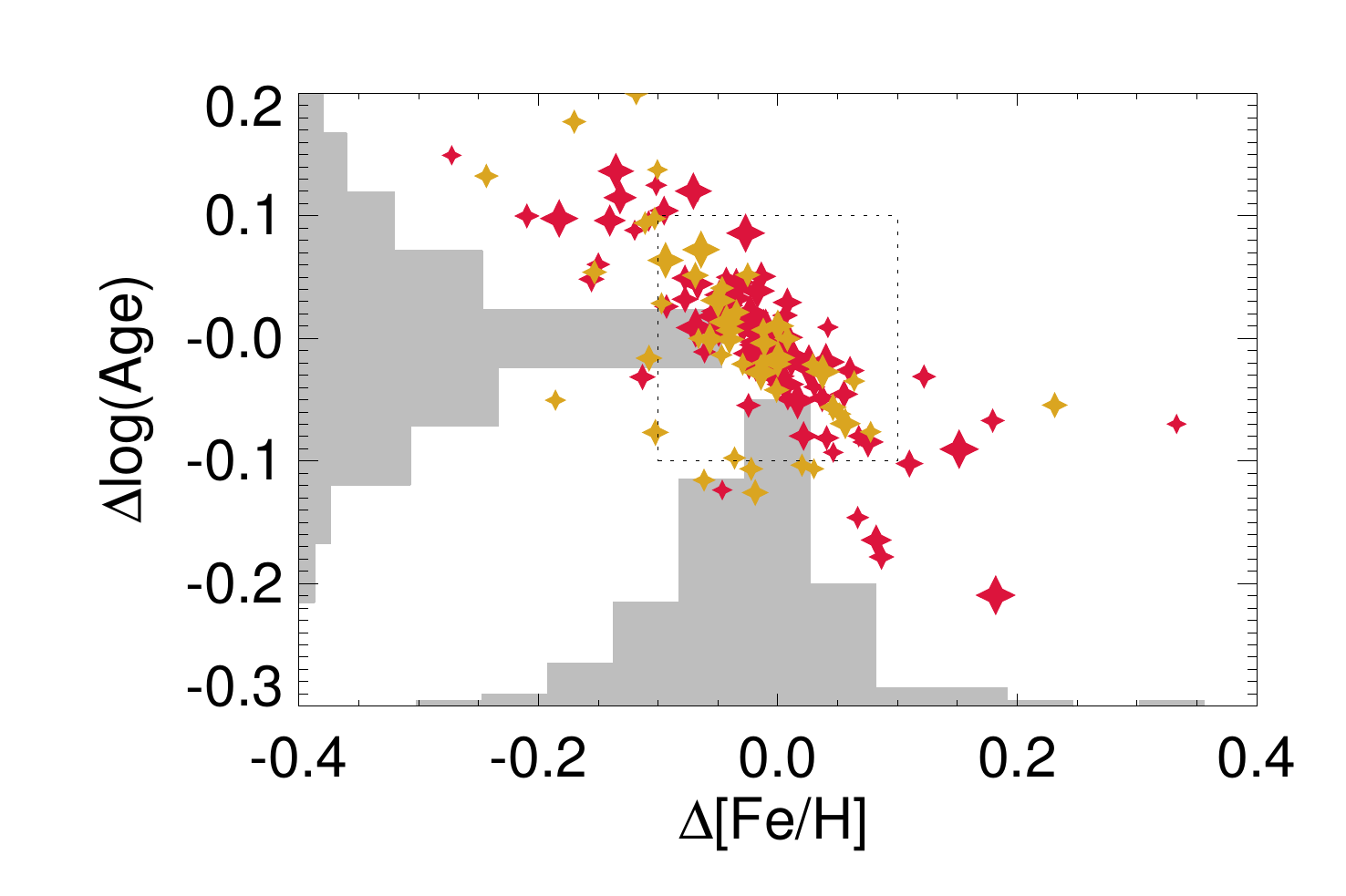}
     \includegraphics[width=0.49\textwidth]{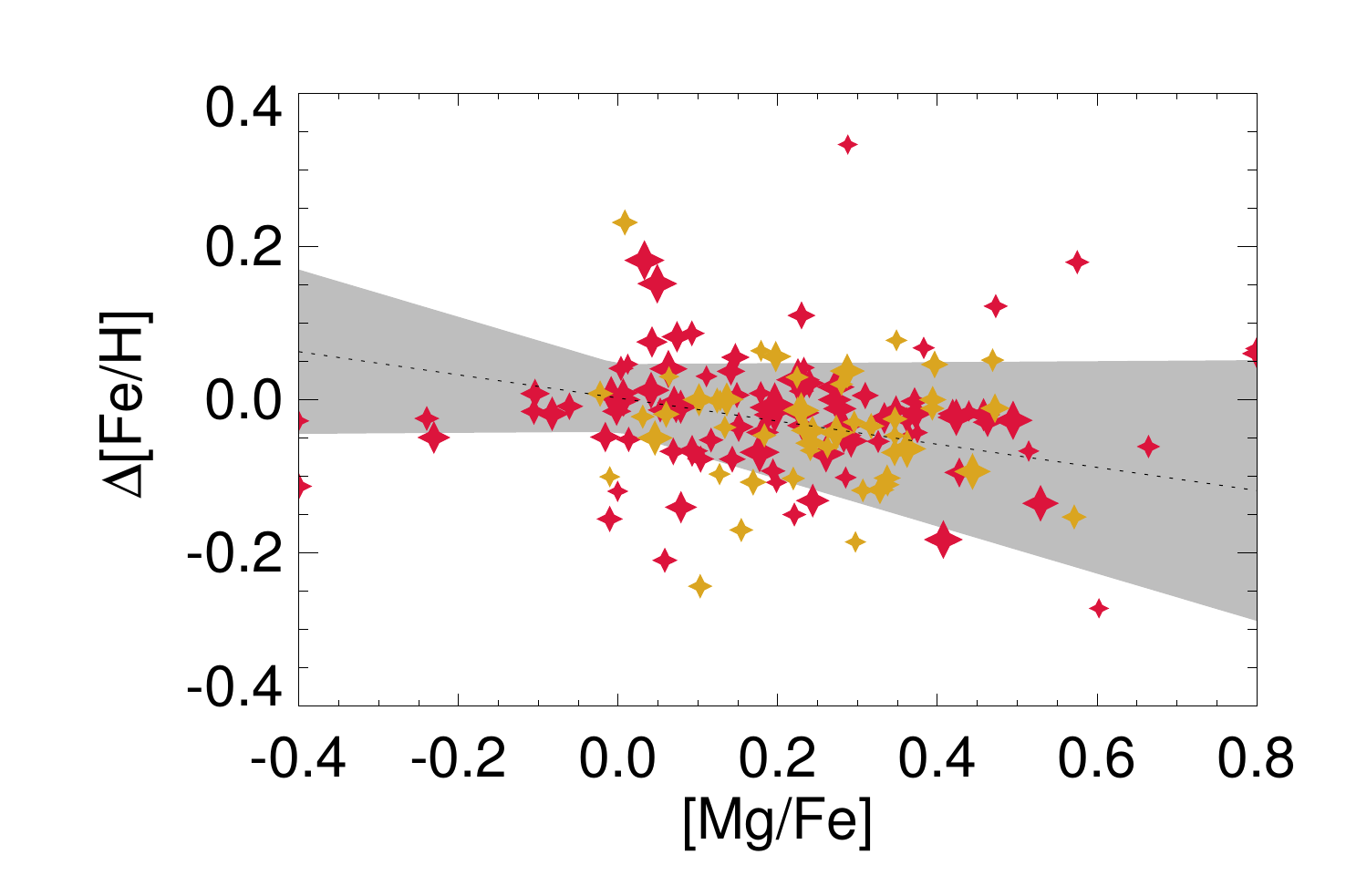}
    \caption{Comparison of the age and [Fe/H] measurements using the models in Paper I (same models but without $\alpha$ enhancement) and the models used here. The left panel shows the differences in the measured age and [Fe/H] with underlying histograms in gray color. The color (yellow and red) represents the $z\sim0.39$ and the $z\sim0.55$ sample, respectively. The dotted box delineates uncertainties of 0.1 dex. The right panel shows the difference in [Fe/H] with the best linear fit and its uncertainty (dotted line and gray shading). The uncertainties in [Fe/H] and age lie along the age-metallicity degeneracy direction with a slight dependence on [Mg/Fe].}
    \label{fig:oldnew}
\end{figure}

Figure \ref{fig:oldnew} shows the comparisons of the results from both methods. The $x$- and $y$- axes represent the results from this paper subtracted from the results using the method in Paper I\@. The left panel shows that the differences in [Fe/H] and age lies along the age--metallicity degeneracy, with means of $-0.01\pm0.01$ dex and $0.00\pm0.01$ dex, respectively. The mode of the distributions are zero. I.e., there are no significant systematic differences in the [Fe/H] nor age measured by either method. Most of the discrepancies are within 0.1 dex in either parameter, as shown by the dotted box. However, the distribution of the differences in measured [Fe/H] ($\Delta$[Fe/H]) is slightly asymmetric, with a heavier distribution toward negative values. This means that we are more likely to have underestimated [Fe/H] in this paper compared to Paper I\@.

The right panel shows $\Delta$[Fe/H] vs.\ [Mg/Fe] along with the best-fit line. Although the best-fit slope is statistically consistent with zero, its best-fit value suggests that when we [Mg/Fe] is high, we possibly underestimate [Fe/H] with the method in Paper I. The reason for this is also likely due to the response function of Fe and Mg. The spectrum in the 4000--4400 \angstrom\ range responds to Mg and Fe in the opposite direction. Therefore, the models without elemental enhancements are likely to underestimate the [Fe/H] when fit to a Mg-enhanced spectrum.

\section{Degeneracies between age, [Fe/H] and [Mg/Fe]}
\label{appendix:uncertainties}
\begin{figure}
    \centering
    \includegraphics[width=0.32\textwidth]{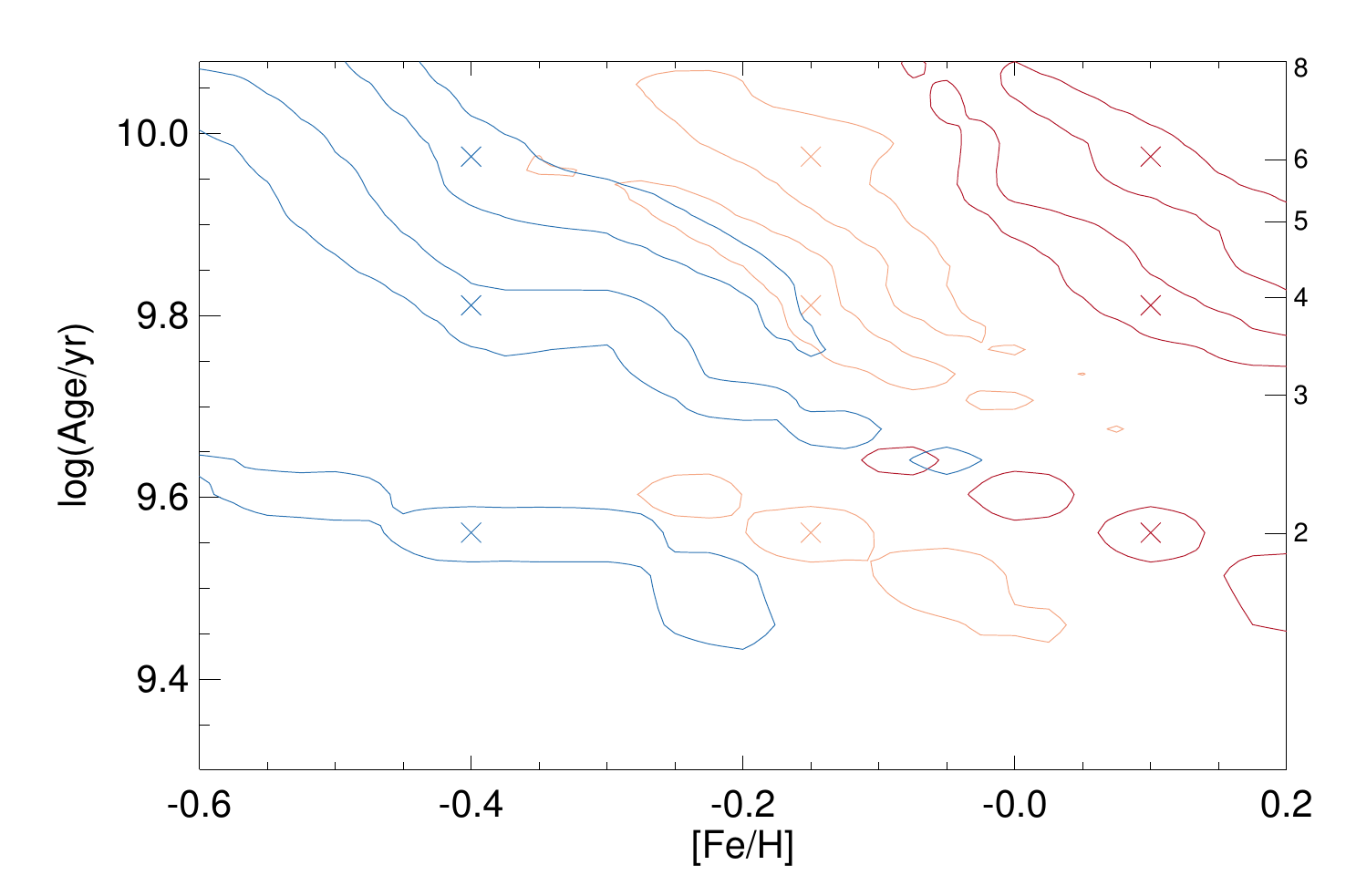}
    \includegraphics[width=0.32\textwidth]{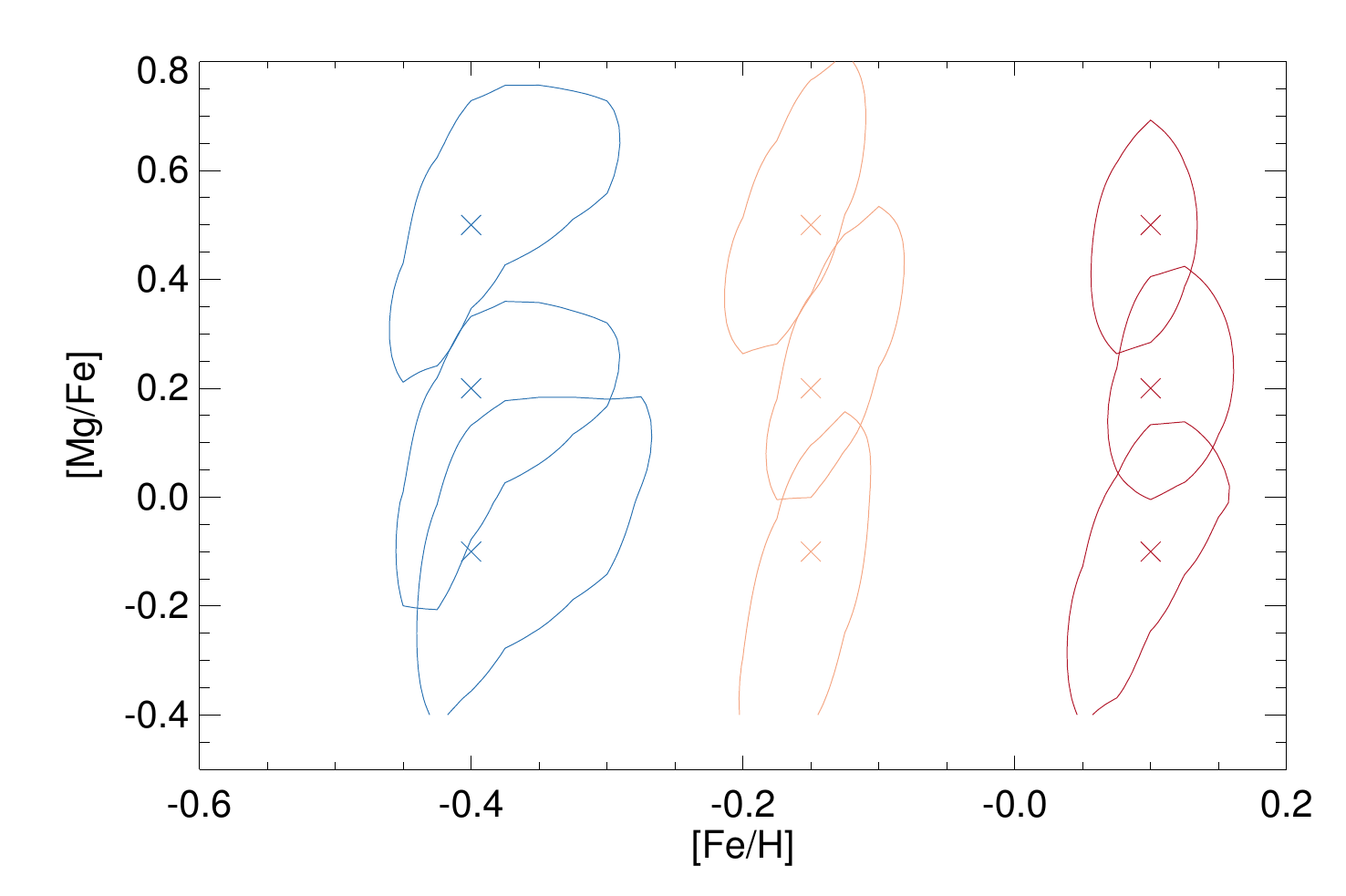}
    \includegraphics[width=0.32\textwidth]{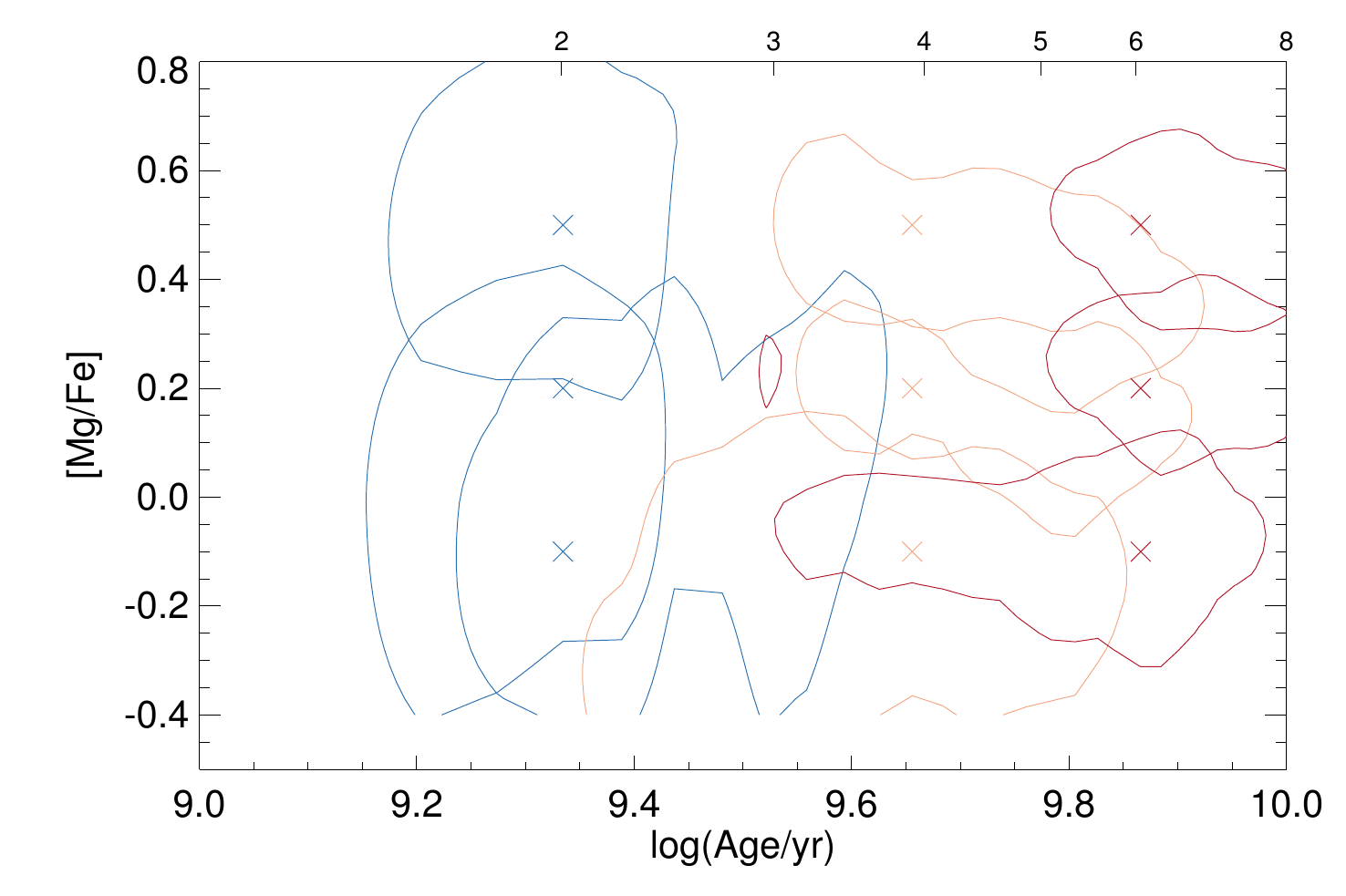}\\
    \includegraphics[width=0.32\textwidth]{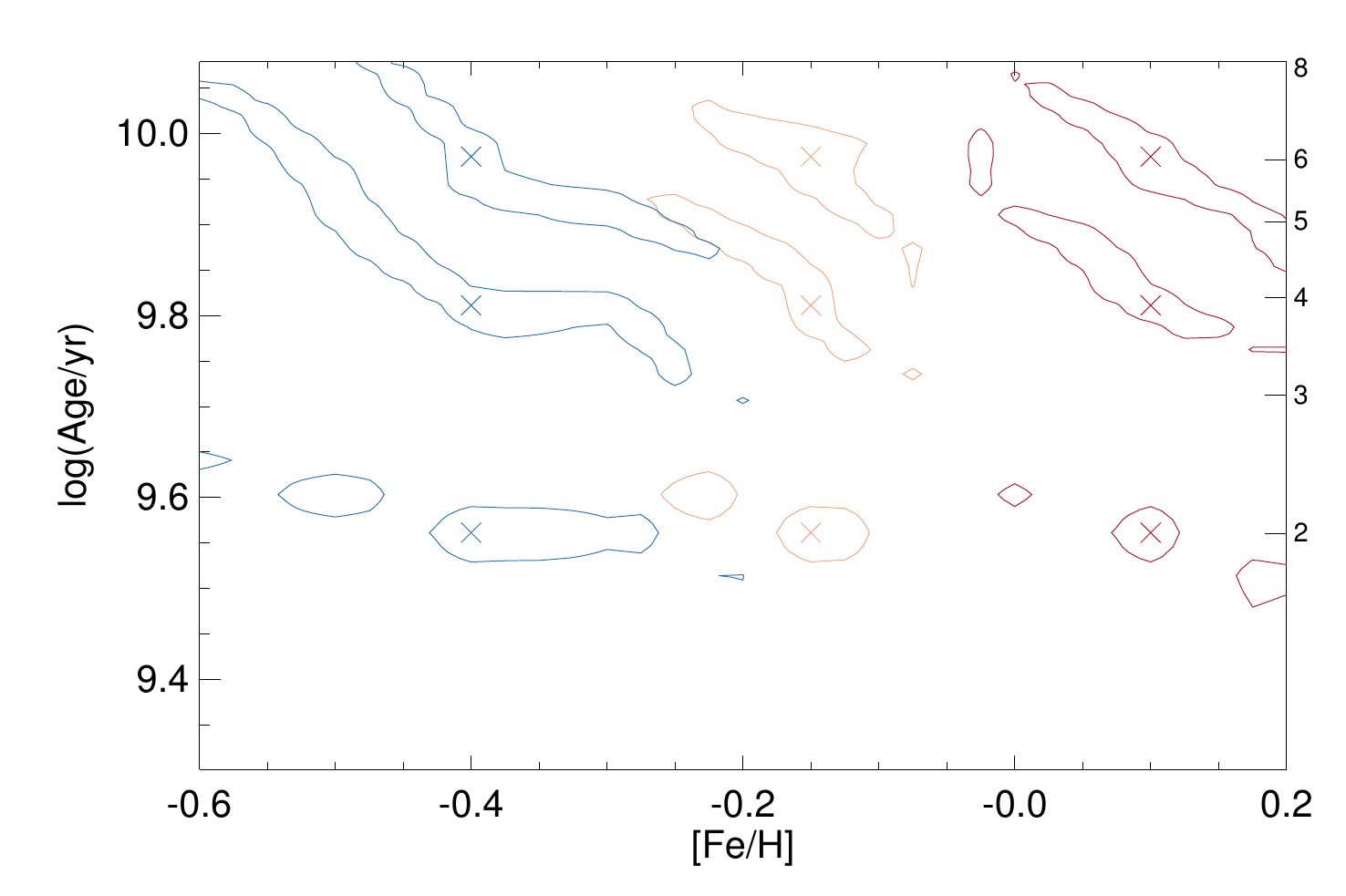}
    \includegraphics[width=0.32\textwidth]{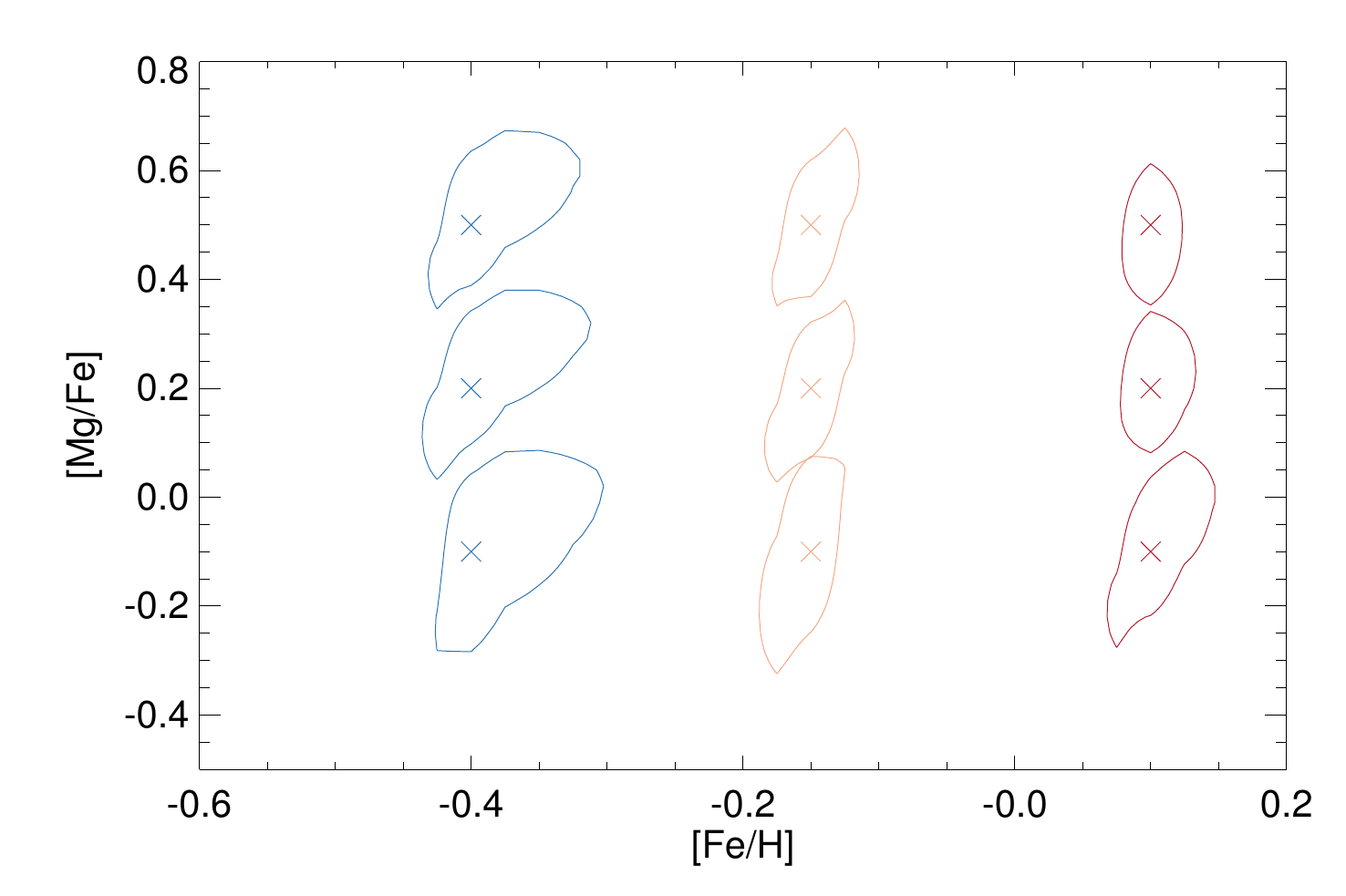}
    \includegraphics[width=0.32\textwidth]{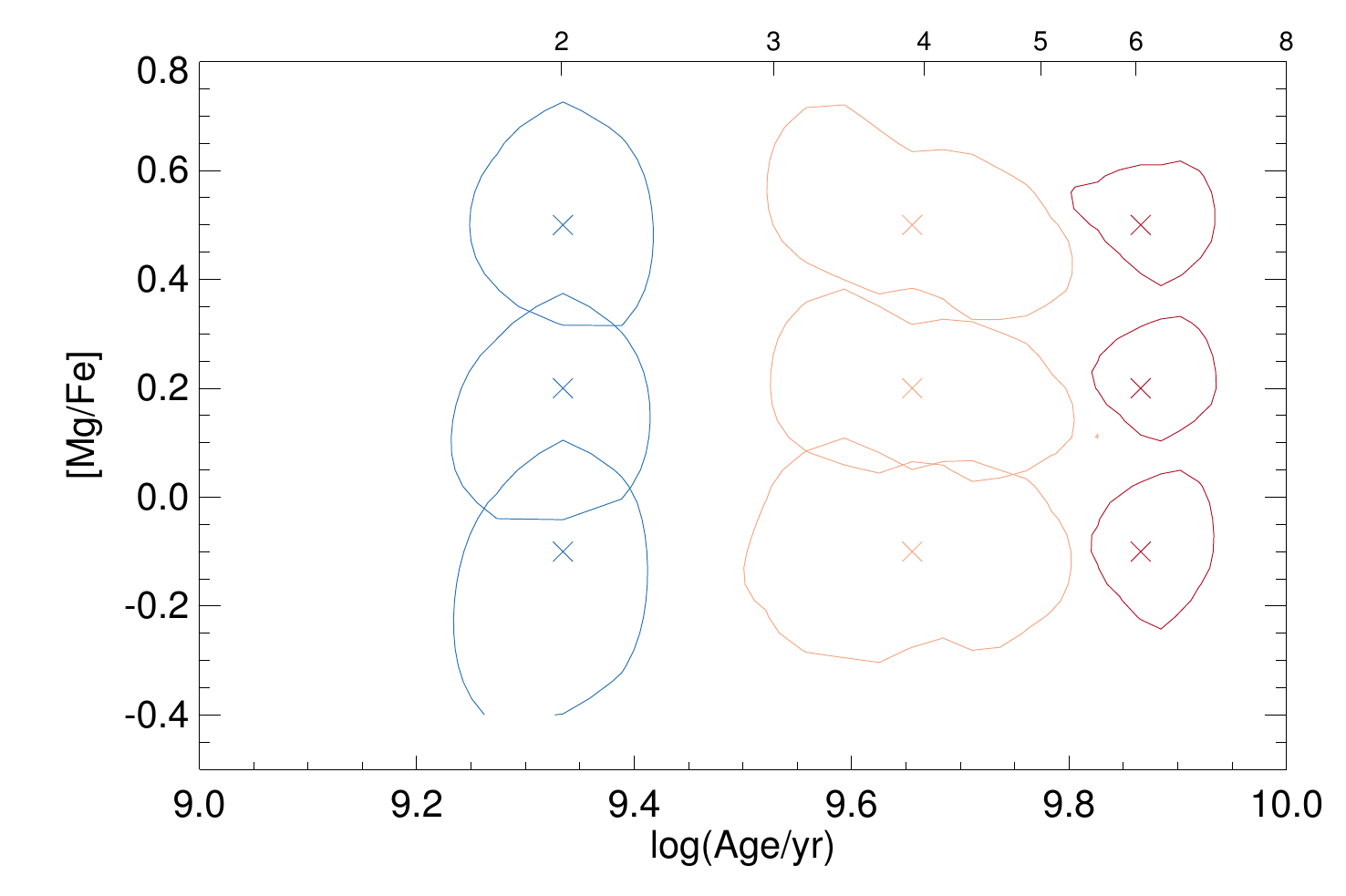}\\
    \caption{Uncertainties in the measured ages, [Fe/H]'s, and [Mg/Fe]'s according to degeneracies between the three parameters. Each contour shows
the 1$\sigma$ range in uncertainties of the x-y parameters, while the third parameter is fixed at the correct value (at [Mg/Fe]$= 0.1$ dex, Age$=3$ Gyr, and [Fe/H]$=-0.15$ dex from left to right). The true parameters of the x-y axis are shown as cross marks. The S/N of each spectrum is 10 \angstrom$^{-1}$ (upper panels) and 25 \angstrom$^{-1}$ (lower panels).}
    \label{fig:worm}
\end{figure}

In this section, we explore the systematic uncertainties based on the degeneracies between the three parameters of interest: age, [Fe/H] and [Mg/Fe]. To do so, we select a set of reference spectra from the same noiseless SSP grid generated in Section \ref{sec:modelfitting}.  Then, we add Gaussian noise to reach S/N of 10 \angstrom$^{-1}$ (lower S/N limit of our data) and 25 \angstrom$^{-1}$. We then calculate their $\chi^2$s by comparing them to the the noiseless SSP grid. The results are shown in Figure \ref{fig:worm}. Each figure shows contours of $1\sigma$ uncertainty based on the 2-D posterior probability distribution functions of the degeneracy between the two parameters on the x-y axis. The third parameter is fixed at [Mg/Fe]$= 0.1$ dex, Age$=3$ Gyr, and [Fe/H]$=-0.15$ dex (from left to right respectively).

Based on Figure \ref{fig:worm}, the correlation between uncertainties of the measured parameters is strongest for [Fe/H] and age (left column). This is known as the age-metallicity degeneracy. When the [Fe/H] is underestimated, the age is overestimated (and vice versa). In the middle column, we observe a slight positive correlation between the uncertainties of [Mg/Fe] and of [Fe/H]. If [Mg/Fe] is overestimated, [Fe/H] will likely also be overestimated. This is due to the overlapping response functions of both elements in the the 4000-4400 \angstrom\ range. This is consistent with what we found in Appendix \ref{appendix:paper1compare}. Lastly, as shown in the left column, the uncertainties in [Mg/Fe] and age are the most radially symmetrical. This suggests that [Mg/Fe] and age have the least degeneracy.    

The dominant source of the systematic uncertainty in [Fe/H] is the degeneracy between [Fe/H] and age (left column). This degeneracy contributes $\sim0.2$ dex to the total uncertainty of [Fe/H] at S/N$=10$ \angstrom$^{-1}$ (this reduces to $\sim0.1$ dex at S/N$=25$ \angstrom$^{-1}$). The uncertainty in [Fe/H] also depends on the [Fe/H] value. It is larger at lower [Fe/H]'s and can be as large as 0.3 dex at low S/N. The degeneracy between [Fe/H] and [Mg/Fe] (middle column) is a smaller source of [Fe/H] uncertainty, contributing less than half of the uncertainty that is generated by the degeneracy between [Fe/H] and age (left column). As for the uncertainty in age, the dominant source is also the degeneracy between [Fe/H] and age. However, at low [Mg/Fe] values (less than solar), the systematic uncertainty in age increases significantly. Nevertheless, we need not be too concerned because most of the quiescent galaxies are [Mg/Fe] enhanced. Finally, the uncertainty in [Mg/Fe] comes equally from the degeneracy of [Mg/Fe] with age and with [Fe/H]. Both contribute $\sim$ 0.2-0.3 dex at low S/N and $\sim0.1$ dex at high S/N.

\section{The impact of enriched outflow and enriched inflow on measured mass-loading factors}
\label{appendix:enrichedoutflow}
 In this section, we separately investigate the effect of the assumed pristine inflow and the assumed gas-phase metallicity of the outflow on our measured mass-loading factors. We first explore the case of an enriched inflow. If the infalling gas has a metallicity of $Z_\text{inf}$, we can modify Equation \ref{equation:chem1} to 
\begin{equation}
    dM_{Z,g} = ydM_*-Z_gdM_*-\frac{\eta}{1-R}Z_gdM_*+Z_\text{inf}  dM_{\text{inf}}
\label{equation:chem1_withZinflow}
\end{equation}
where $dM_{\text{inf}}$ is the mass inflow rate. We can substitute $Z_gdM_*$ with $dM_{Z,*}$, integrate the equation, and divide all terms by $dM_*$ to get
\begin{equation}
    Z_gr_g = y-(1+\frac{\eta}{1-R})Z_*+<Z_\text{inf}>\frac{M_\text{inf}}{M_*}
\end{equation}
Assuming the galaxy started with no gas, we can estimate the total infall mass for quiescent galaxies as the current stellar mass plus the total outflow mass: $M_\text{inf} = (1+\frac{\eta}{1-R})M_*$. The equation for quiescent galaxies is then
\begin{equation}
    Z_{*} \approx \frac{y}{1+\frac{<\eta>}{1-R}}+<Z_\text{inf}> \quad \text{or} \quad <\eta> \approx \bigg(\frac{y}{Z_*-<Z_\text{inf}>}-1\bigg)(1-R)
\end{equation}
 
\begin{figure}
     \centering
     \includegraphics[width=0.5\textwidth]{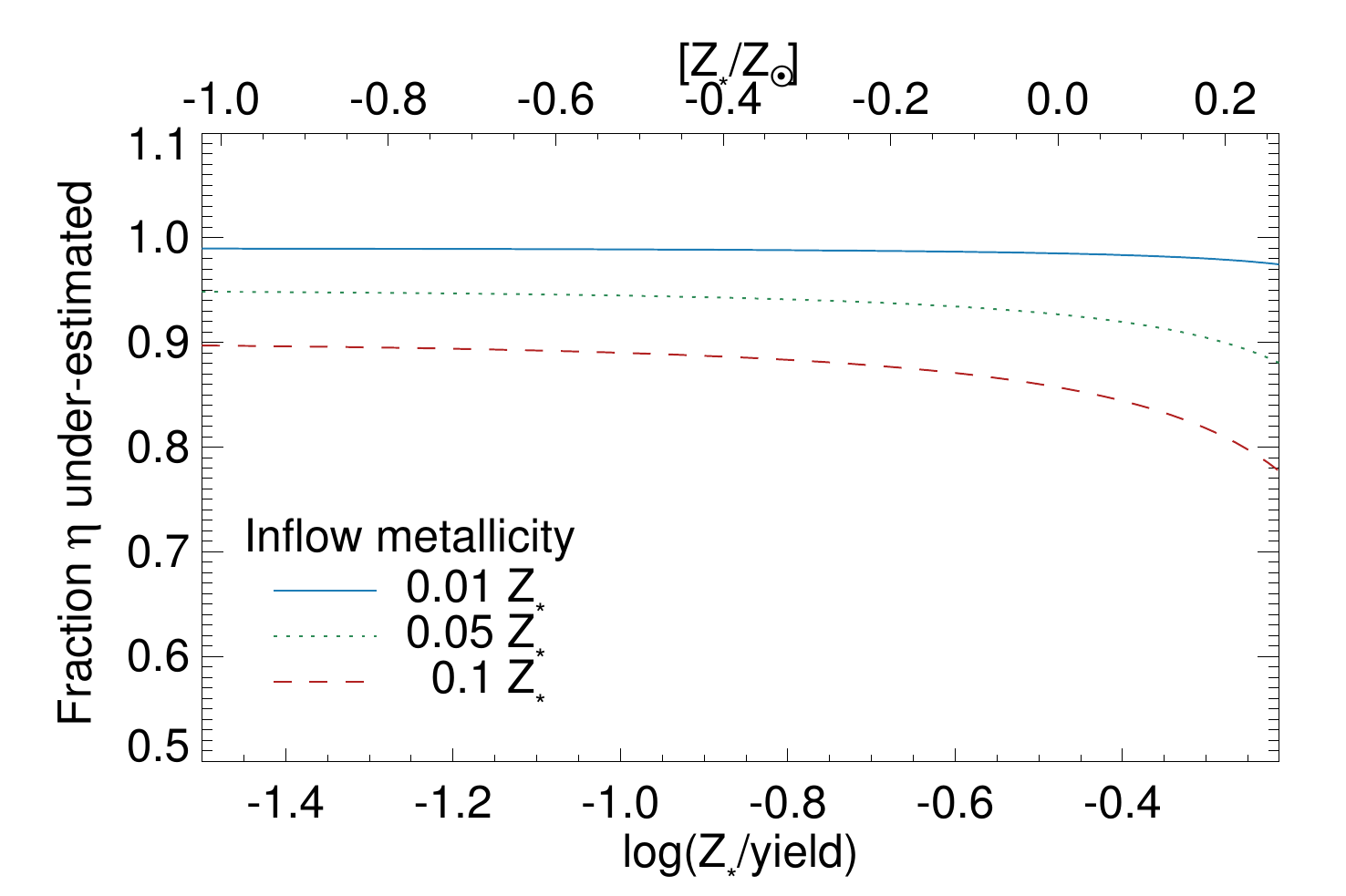}
     \caption{The underestimation of the mass-loading factor (the ratio between $\eta$ derived assuming pristine infalling gas to $\eta$ derived with enriched inflow) as a function of stellar metallicity. The bottom $x$-axis is the metallicity in units of the yield. The top $x$-axis shows the metallicity when the yield is assumed to be $3Z_\odot$. Each line corresponds to a different assumed metallicity of the inflow.} 
     \label{fig:nta_enrich}
 \end{figure}

 The equation above suggests that we would underestimate the mass-loading factor by assuming that the inflowing gas in pristine. The amount of the underestimation is shown in Figure \ref{fig:nta_enrich}. The $y$-axis is the ratio between the $\eta$ derived assuming pristine inflow to the $\eta$ derived from the model with enriched inflow. The amount of the underestimation depends on both inflow and stellar metallicity. For simplicity in quantifying the amount of the underestimation, we parameterize the inflow metallicity as a fixed fraction of the final stellar metallicity: $Z_\text{inf} = \alpha Z_*$. (This is similar to the parameterization used by \citealt{FinlatorDave2008} but with the stellar metallicity instead of the gas-phase metallicity.) 
 
The underestimation of the mass-loading factor when gas inflow is assumed to be pristine is estimated to be at most $20\%$ on a linear scale. Studies of infalling \ion{H}{1} absorbing complexes generally have found that \ion{H}{1} filaments and the circum-galactic medium are metal-poor, $\log Z/Z_\odot < -1$ \citep[e.g.,][]{Churchill2012,Hafen2017}. Thus, based on Figure \ref{fig:nta_enrich}, even in a very enriched inflow case at the high-mass end ($Z_\text{inf}=0.1Z_*\sim -1$ dex), the estimated $\eta$ would be at most underestimated by 20\%, or less than 0.1 dex.

 Now, we investigate the assumption of the perfectly mixed ISM, i.e., the outflow metallicity is the same as the ISM metallicity. There is observational evidence that the outflows might be more metal-rich than the ISM. For example, \citet{Chisholm2018} measured the outflow metallicities of five nearby star-forming galaxies based on rest-frame ultraviolet absorption lines. The authors found that on average, the outflow metallicities are 2.6 times larger than the ISM with no significant dependence on galaxy mass. 
 
 If the outflow is more enriched than the ISM, we will overestimate the mass-loading factors by assuming that the outflow metallicity is the same as the ISM metallicity. Let's assume that the outflow metallicity is $\beta Z_g$ instead of $Z_g$, we can modify Equation \ref{equation:chem1} to 
 \begin{equation}
    dM_{Z,g} = ydM_*-Z_gdM_*-\frac{\eta}{1-R}\beta Z_gdM_*
\label{equation:chem1_withZoutflow}
\end{equation}
This means that our estimated $\eta$ in this paper are actually $\beta\eta$. Therefore, if $\beta$ is 2.6 then we would overestimate the mass-loading factors by the same amount, which is 0.4 dex in log scale. If $\beta$ is really independent of mass, as found in \citet{Chisholm2018}, this should have no effect on the derived power-law index of the relation between $\eta$ and $M_*$.

\begin{thebibliography}{}
\expandafter\ifx\csname natexlab\endcsname\relax\def\natexlab#1{#1}\fi
\providecommand{\url}[1]{\href{#1}{#1}}

\bibitem[{{Asplund} {et~al.}(2009){Asplund}, {Grevesse}, {Sauval}, \&
  {Scott}}]{Asplund2009}
{Asplund}, M., {Grevesse}, N., {Sauval}, A.~J., \& {Scott}, P. 2009, Annual
  Review of Astronomy and Astrophysics, 47, 481

\bibitem[{{Bah{\'e}} \& {McCarthy}(2015)}]{Bahe2015}
{Bah{\'e}}, Y.~M., \& {McCarthy}, I.~G. 2015, \mnras, 447, 969

\bibitem[{{Belfiore} {et~al.}(2017){Belfiore}, {Maiolino}, {Maraston},
  {Emsellem}, {Bershady}, {Masters}, {Bizyaev}, {Boquien}, {Brownstein},
  {Bundy}, {Diamond-Stanic}, {Drory}, {Heckman}, {Law}, {Malanushenko},
  {Oravetz}, {Pan}, {Roman-Lopes}, {Thomas}, {Weijmans}, {Westfall}, \&
  {Yan}}]{Belfiore2017}
{Belfiore}, F., {Maiolino}, R., {Maraston}, C., {et~al.} 2017, \mnras, 466,
  2570

\bibitem[{{Belfiore} {et~al.}(2019){Belfiore}, {Westfall}, {Schaefer},
  {Cappellari}, {Ji}, {Bershady}, {Tremonti}, {Law}, {Yan}, {Bundy}, {Shetty},
  {Drory}, {Thomas}, {Emsellem}, \& {S{\'a}nchez}}]{Belfiore2019}
{Belfiore}, F., {Westfall}, K.~B., {Schaefer}, A., {et~al.} 2019, arXiv
  e-prints, arXiv:1901.00866

\bibitem[{{Blanton} \& {Roweis}(2007)}]{Blanton2007}
{Blanton}, M.~R., \& {Roweis}, S. 2007, \aj, 133, 734

\bibitem[{{Bolatto} {et~al.}(2013){Bolatto}, {Warren}, {Leroy}, {Walter},
  {Veilleux}, {Ostriker}, {Ott}, {Zwaan}, {Fisher}, {Weiss}, {Rosolowsky}, \&
  {Hodge}}]{Bolatto2013}
{Bolatto}, A.~D., {Warren}, S.~R., {Leroy}, A.~K., {et~al.} 2013, \nat, 499,
  450

\bibitem[{{Boselli} {et~al.}(2014){Boselli}, {Cortese}, {Boquien}, {Boissier},
  {Catinella}, {Lagos}, \& {Saintonge}}]{Boselli2014}
{Boselli}, A., {Cortese}, L., {Boquien}, M., {et~al.} 2014, \aap, 564, A66

\bibitem[{{Bouch{\'e}} {et~al.}(2012){Bouch{\'e}}, {Hohensee}, {Vargas},
  {Kacprzak}, {Martin}, {Cooke}, \& {Churchill}}]{Bouche2012}
{Bouch{\'e}}, N., {Hohensee}, W., {Vargas}, R., {et~al.} 2012, \mnras, 426, 801

\bibitem[{{Bundy} {et~al.}(2015){Bundy}, {Bershady}, {Law}, {Yan}, {Drory},
  {MacDonald}, {Wake}, {Cherinka}, {S{\'a}nchez-Gallego}, {Weijmans}, {Thomas},
  {Tremonti}, {Masters}, {Coccato}, {Diamond-Stanic}, {Arag{\'o}n-Salamanca},
  {Avila-Reese}, {Badenes}, {Falc{\'o}n-Barroso}, {Belfiore}, {Bizyaev},
  {Blanc}, {Bland-Hawthorn}, {Blanton}, {Brownstein}, {Byler}, {Cappellari},
  {Conroy}, {Dutton}, {Emsellem}, {Etherington}, {Frinchaboy}, {Fu}, {Gunn},
  {Harding}, {Johnston}, {Kauffmann}, {Kinemuchi}, {Klaene}, {Knapen},
  {Leauthaud}, {Li}, {Lin}, {Maiolino}, {Malanushenko}, {Malanushenko}, {Mao},
  {Maraston}, {McDermid}, {Merrifield}, {Nichol}, {Oravetz}, {Pan}, {Parejko},
  {Sanchez}, {Schlegel}, {Simmons}, {Steele}, {Steinmetz}, {Thanjavur},
  {Thompson}, {Tinker}, {van den Bosch}, {Westfall}, {Wilkinson}, {Wright},
  {Xiao}, \& {Zhang}}]{Bundy2015}
{Bundy}, K., {Bershady}, M.~A., {Law}, D.~R., {et~al.} 2015, \apj, 798, 7

\bibitem[{{Chisholm} {et~al.}(2018){Chisholm}, {Tremonti}, \&
  {Leitherer}}]{Chisholm2018}
{Chisholm}, J., {Tremonti}, C., \& {Leitherer}, C. 2018, \mnras, 481, 1690

\bibitem[{{Chisholm} {et~al.}(2017){Chisholm}, {Tremonti}, {Leitherer}, \&
  {Chen}}]{Chisholm2017}
{Chisholm}, J., {Tremonti}, C.~A., {Leitherer}, C., \& {Chen}, Y. 2017, \mnras,
  469, 4831

\bibitem[{{Choi} {et~al.}(2014){Choi}, {Conroy}, {Moustakas}, {Graves},
  {Holden}, {Brodwin}, {Brown}, \& {van Dokkum}}]{Choi2014}
{Choi}, J., {Conroy}, C., {Moustakas}, J., {et~al.} 2014, \apj, 792, 95

\bibitem[{{Choi} {et~al.}(2016){Choi}, {Dotter}, {Conroy}, {Cantiello},
  {Paxton}, \& {Johnson}}]{Choi2016}
{Choi}, J., {Dotter}, A., {Conroy}, C., {et~al.} 2016, \apj, 823, 102

\bibitem[{{Churchill} {et~al.}(2012){Churchill}, {Kacprzak}, {Steidel},
  {Spitler}, {Holtzman}, {Nielsen}, \& {Trujillo-Gomez}}]{Churchill2012}
{Churchill}, C.~W., {Kacprzak}, G.~G., {Steidel}, C.~C., {et~al.} 2012, \apj,
  760, 68

\bibitem[{{Cicone} {et~al.}(2014){Cicone}, {Maiolino}, {Sturm},
  {Graci{\'a}-Carpio}, {Feruglio}, {Neri}, {Aalto}, {Davies}, {Fiore},
  {Fischer}, {Garc{\'\i}a-Burillo}, {Gonz{\'a}lez-Alfonso}, {Hailey-Dunsheath},
  {Piconcelli}, \& {Veilleux}}]{Cicone2014}
{Cicone}, C., {Maiolino}, R., {Sturm}, E., {et~al.} 2014, \aap, 562, A21

\bibitem[{{Cid Fernandes}(2018)}]{CidFernandes2018}
{Cid Fernandes}, R. 2018, \mnras, 480, 4480

\bibitem[{{Concas} {et~al.}(2017){Concas}, {Popesso}, {Brusa}, {Mainieri},
  {Erfanianfar}, \& {Morselli}}]{Concas2017}
{Concas}, A., {Popesso}, P., {Brusa}, M., {et~al.} 2017, \aap, 606, A36

\bibitem[{{Conroy} {et~al.}(2014){Conroy}, {Graves}, \& {van
  Dokkum}}]{Conroy2014}
{Conroy}, C., {Graves}, G.~J., \& {van Dokkum}, P.~G. 2014, \apj, 780, 33

\bibitem[{{Conroy} {et~al.}(2009){Conroy}, {Gunn}, \& {White}}]{Conroy2009}
{Conroy}, C., {Gunn}, J.~E., \& {White}, M. 2009, \apj, 699, 486

\bibitem[{{Conroy} {et~al.}(2018){Conroy}, {Villaume}, {van Dokkum}, \&
  {Lind}}]{Conroy2018}
{Conroy}, C., {Villaume}, A., {van Dokkum}, P.~G., \& {Lind}, K. 2018, \apj,
  854, 139

\bibitem[{{Cooper} {et~al.}(2008){Cooper}, {Tremonti}, {Newman}, \&
  {Zabludoff}}]{Cooper2008}
{Cooper}, M.~C., {Tremonti}, C.~A., {Newman}, J.~A., \& {Zabludoff}, A.~I.
  2008, \mnras, 390, 245

\bibitem[{{Dav{\'e}} {et~al.}(2011){Dav{\'e}}, {Finlator}, \&
  {Oppenheimer}}]{Dave2011}
{Dav{\'e}}, R., {Finlator}, K., \& {Oppenheimer}, B.~D. 2011, \mnras, 416, 1354

\bibitem[{{Dekel} \& {Silk}(1986)}]{DekelSilk1986}
{Dekel}, A., \& {Silk}, J. 1986, \apj, 303, 39

\bibitem[{{Estrada-Carpenter} {et~al.}(2019){Estrada-Carpenter}, {Papovich},
  {Momcheva}, {Brammer}, {Long}, {Quadri}, {Bridge}, {Dickinson}, {Ferguson},
  {Finkelstein}, {Giavalisco}, {Gosmeyer}, {Lotz}, {Salmon}, {Skelton},
  {Trump}, \& {Weiner}}]{Estrada-Carpenter2019}
{Estrada-Carpenter}, V., {Papovich}, C., {Momcheva}, I., {et~al.} 2019, \apj,
  870, 133

\bibitem[{{Faber} {et~al.}(2003){Faber}, {Phillips}, {Kibrick}, {Alcott},
  {Allen}, {Burrous}, {Cantrall}, {Clarke}, {Coil}, {Cowley}, {Davis}, {Deich},
  {Dietsch}, {Gilmore}, {Harper}, {Hilyard}, {Lewis}, {McVeigh}, {Newman},
  {Osborne}, {Schiavon}, {Stover}, {Tucker}, {Wallace}, {Wei}, {Wirth}, \&
  {Wright}}]{Faber2003}
{Faber}, S.~M., {Phillips}, A.~C., {Kibrick}, R.~I., {et~al.} 2003, in Society
  of Photo-Optical Instrumentation Engineers (SPIE) Conference Series, Vol.
  4841, Instrument Design and Performance for Optical/Infrared Ground-based
  Telescopes, ed. M.~{Iye} \& A.~F.~M. {Moorwood}, 1657--1669

\bibitem[{{Finlator} \& {Dav{\'e}}(2008)}]{FinlatorDave2008}
{Finlator}, K., \& {Dav{\'e}}, R. 2008, \mnras, 385, 2181

\bibitem[{{Fitzpatrick} \& {Graves}(2015)}]{Fitzpatrick2015}
{Fitzpatrick}, P.~J., \& {Graves}, G.~J. 2015, \mnras, 447, 1383

\bibitem[{{Gallazzi} {et~al.}(2014){Gallazzi}, {Bell}, {Zibetti}, {Brinchmann},
  \& {Kelson}}]{Gallazzi2014}
{Gallazzi}, A., {Bell}, E.~F., {Zibetti}, S., {Brinchmann}, J., \& {Kelson},
  D.~D. 2014, \apj, 788, 72

\bibitem[{{Gallazzi} {et~al.}(2005){Gallazzi}, {Charlot}, {Brinchmann},
  {White}, \& {Tremonti}}]{Gallazzi2005}
{Gallazzi}, A., {Charlot}, S., {Brinchmann}, J., {White}, S. D.~M., \&
  {Tremonti}, C.~A. 2005, \mnras, 362, 41

\bibitem[{{Garnett}(2002)}]{Garnett2002}
{Garnett}, D.~R. 2002, \apj, 581, 1019

\bibitem[{{Garnett} \& {Shields}(1987)}]{Garnett1987}
{Garnett}, D.~R., \& {Shields}, G.~A. 1987, \apj, 317, 82

\bibitem[{{Ge} {et~al.}(2018){Ge}, {Yan}, {Cappellari}, {Mao}, {Li}, \&
  {Lu}}]{Ge2018}
{Ge}, J., {Yan}, R., {Cappellari}, M., {et~al.} 2018, \mnras, 478, 2633

\bibitem[{{Gobat} {et~al.}(2018){Gobat}, {Daddi}, {Magdis}, {Bournaud},
  {Sargent}, {Martig}, {Jin}, {Finoguenov}, {B{\'e}thermin}, {Hwang},
  {Renzini}, {Wilson}, {Aretxaga}, {Yun}, {Strazzullo}, \&
  {Valentino}}]{Gobat2018}
{Gobat}, R., {Daddi}, E., {Magdis}, G., {et~al.} 2018, Nature Astronomy, 2, 239

\bibitem[{{Greene} {et~al.}(2012){Greene}, {Murphy}, {Comerford}, {Gebhardt},
  \& {Adams}}]{Greene2012}
{Greene}, J.~E., {Murphy}, J.~D., {Comerford}, J.~M., {Gebhardt}, K., \&
  {Adams}, J.~J. 2012, \apj, 750, 32

\bibitem[{{Greene} {et~al.}(2013){Greene}, {Murphy}, {Graves}, {Gunn},
  {Raskutti}, {Comerford}, \& {Gebhardt}}]{Greene2013}
{Greene}, J.~E., {Murphy}, J.~D., {Graves}, G.~J., {et~al.} 2013, \apj, 776, 64

\bibitem[{{Griffith} {et~al.}(2019){Griffith}, {Martini}, \&
  {Conroy}}]{Griffith2019}
{Griffith}, E., {Martini}, P., \& {Conroy}, C. 2019, \mnras, 484, 562

\bibitem[{{Guo} {et~al.}(2016){Guo}, {Gonzalez-Perez}, {Guo}, {Schaller},
  {Furlong}, {Bower}, {Cole}, {Crain}, {Frenk}, {Helly}, {Lacey}, {Lagos},
  {Mitchell}, {Schaye}, \& {Theuns}}]{Guo2016}
{Guo}, Q., {Gonzalez-Perez}, V., {Guo}, Q., {et~al.} 2016, \mnras, 461, 3457

\bibitem[{{Hafen} {et~al.}(2017){Hafen}, {Faucher-Gigu{\`e}re},
  {Angl{\'e}s-Alc{\'a}zar}, {Kere{\v{s}}}, {Feldmann}, {Chan}, {Quataert},
  {Murray}, \& {Hopkins}}]{Hafen2017}
{Hafen}, Z., {Faucher-Gigu{\`e}re}, C.-A., {Angl{\'e}s-Alc{\'a}zar}, D.,
  {et~al.} 2017, \mnras, 469, 2292

\bibitem[{{Harrison} {et~al.}(2011){Harrison}, {Colless}, {Kuntschner},
  {Couch}, {de Propris}, \& {Pracy}}]{Harrison2011}
{Harrison}, C.~D., {Colless}, M., {Kuntschner}, H., {et~al.} 2011, \mnras, 413,
  1036

\bibitem[{{Hayward} \& {Hopkins}(2017)}]{Hayward2017}
{Hayward}, C.~C., \& {Hopkins}, P.~F. 2017, \mnras, 465, 1682

\bibitem[{{J{\o}rgensen} {et~al.}(2018){J{\o}rgensen}, {Chiboucas}, {Webb}, \&
  {Woodrum}}]{Jorgensen2018}
{J{\o}rgensen}, I., {Chiboucas}, K., {Webb}, K., \& {Woodrum}, C. 2018, \aj,
  156, 224

\bibitem[{{Kacprzak} {et~al.}(2015){Kacprzak}, {Yuan}, {Nanayakkara},
  {Kobayashi}, {Tran}, {Kewley}, {Glazebrook}, {Spitler}, {Taylor}, {Cowley},
  {Labbe}, {Straatman}, \& {Tomczak}}]{Kacprzak2015}
{Kacprzak}, G.~G., {Yuan}, T., {Nanayakkara}, T., {et~al.} 2015, \apj, 802, L26

\bibitem[{{Kenney} \& {Young}(1986)}]{Kenney1986}
{Kenney}, J.~D., \& {Young}, J.~S. 1986, \apjl, 301, L13

\bibitem[{{Kirby} {et~al.}(2013){Kirby}, {Cohen}, {Guhathakurta}, {Cheng},
  {Bullock}, \& {Gallazzi}}]{Kirby2013}
{Kirby}, E.~N., {Cohen}, J.~G., {Guhathakurta}, P., {et~al.} 2013, \apj, 779,
  102

\bibitem[{{Kirby} {et~al.}(2015){Kirby}, {Simon}, \& {Cohen}}]{Kirby2015}
{Kirby}, E.~N., {Simon}, J.~D., \& {Cohen}, J.~G. 2015, \apj, 810, 56

\bibitem[{{Kochanek} {et~al.}(2012){Kochanek}, {Eisenstein}, {Cool},
  {Caldwell}, {Assef}, {Jannuzi}, {Jones}, {Murray}, {Forman}, {Dey}, {Brown},
  {Eisenhardt}, {Gonzalez}, {Green}, \& {Stern}}]{Kochanek2012}
{Kochanek}, C.~S., {Eisenstein}, D.~J., {Cool}, R.~J., {et~al.} 2012, \apjs,
  200, 8

\bibitem[{{Kriek} {et~al.}(2016){Kriek}, {Conroy}, {van Dokkum}, {Shapley},
  {Choi}, {Reddy}, {Siana}, {van de Voort}, {Coil}, \& {Mobasher}}]{Kriek2016}
{Kriek}, M., {Conroy}, C., {van Dokkum}, P.~G., {et~al.} 2016, \nat, 540, 248

\bibitem[{{Kroupa}(2001)}]{Kroupa2001}
{Kroupa}, P. 2001, \mnras, 322, 231

\bibitem[{{Kurucz}(1993)}]{Kurucz1993}
{Kurucz}, R. 1993, SYNTHE Spectrum Synthesis Programs and Line Data. Kurucz
  CD-ROM No. 18. Cambridge, 18

\bibitem[{{Lagos} {et~al.}(2013){Lagos}, {Lacey}, \& {Baugh}}]{Lagos2013}
{Lagos}, C. d.~P., {Lacey}, C.~G., \& {Baugh}, C.~M. 2013, \mnras, 436, 1787

\bibitem[{{Larson}(1974)}]{Larson1974}
{Larson}, R.~B. 1974, \mnras, 169, 229

\bibitem[{{Leethochawalit} {et~al.}(2018){Leethochawalit}, {Kirby}, {Moran},
  {Ellis}, \& {Treu}}]{Leethochawalit2018}
{Leethochawalit}, N., {Kirby}, E.~N., {Moran}, S.~M., {Ellis}, R.~S., \&
  {Treu}, T. 2018, \apj, 856, 15 (Paper I)

\bibitem[{{Lequeux} {et~al.}(1979){Lequeux}, {Peimbert}, {Rayo}, {Serrano}, \&
  {Torres-Peimbert}}]{Lequeux1979}
{Lequeux}, J., {Peimbert}, M., {Rayo}, J.~F., {Serrano}, A., \&
  {Torres-Peimbert}, S. 1979, \aap, 500, 145

\bibitem[{{Lian} {et~al.}(2018){Lian}, {Thomas}, {Maraston}, {Goddard},
  {Comparat}, {Gonzalez-Perez}, \& {Ventura}}]{Lian2018}
{Lian}, J., {Thomas}, D., {Maraston}, C., {et~al.} 2018, \mnras, 474, 1143

\bibitem[{{Lu} {et~al.}(2015){Lu}, {Blanc}, \& {Benson}}]{Lu2015}
{Lu}, Y., {Blanc}, G.~A., \& {Benson}, A. 2015, \apj, 808, 129

\bibitem[{{Lu} {et~al.}(2014){Lu}, {Wechsler}, {Somerville}, {Croton},
  {Porter}, {Primack}, {Behroozi}, {Ferguson}, {Koo}, {Guo}, {Safarzadeh},
  {Finlator}, {Castellano}, {White}, {Sommariva}, \& {Moody}}]{Lu2014}
{Lu}, Y., {Wechsler}, R.~H., {Somerville}, R.~S., {et~al.} 2014, \apj, 795, 123

\bibitem[{{Ma} {et~al.}(2016){Ma}, {Hopkins}, {Faucher-Gigu{\`e}re}, {Zolman},
  {Muratov}, {Kere{\v{s}}}, \& {Quataert}}]{Ma2016}
{Ma}, X., {Hopkins}, P.~F., {Faucher-Gigu{\`e}re}, C.-A., {et~al.} 2016,
  \mnras, 456, 2140

\bibitem[{{Maiolino} {et~al.}(2008){Maiolino}, {Nagao}, {Grazian}, {Cocchia},
  {Marconi}, {Mannucci}, {Cimatti}, {Pipino}, {Ballero}, {Calura}, {Chiappini},
  {Fontana}, {Granato}, {Matteucci}, {Pastorini}, {Pentericci}, {Risaliti},
  {Salvati}, \& {Silva}}]{Maiolino2008}
{Maiolino}, R., {Nagao}, T., {Grazian}, A., {et~al.} 2008, \aap, 488, 463

\bibitem[{{Maiolino} {et~al.}(2012){Maiolino}, {Gallerani}, {Neri}, {Cicone},
  {Ferrara}, {Genzel}, {Lutz}, {Sturm}, {Tacconi}, {Walter}, {Feruglio},
  {Fiore}, \& {Piconcelli}}]{Maiolino2012}
{Maiolino}, R., {Gallerani}, S., {Neri}, R., {et~al.} 2012, \mnras, 425, L66

\bibitem[{{Maoz} {et~al.}(2012){Maoz}, {Mannucci}, \& {Brandt}}]{Maoz2012}
{Maoz}, D., {Mannucci}, F., \& {Brandt}, T.~D. 2012, \mnras, 426, 3282

\bibitem[{{Marigo} \& {Girardi}(2007)}]{Marigo2007}
{Marigo}, P., \& {Girardi}, L. 2007, \aap, 469, 239

\bibitem[{{Markwardt}(2012)}]{Markwardt2012}
{Markwardt}, C. 2012, {MPFIT: Robust non-linear least squares curve fitting},
  Astrophysics Source Code Library, , , ascl:1208.019

\bibitem[{{Mathews} \& {Baker}(1971)}]{MathewsBaker1971}
{Mathews}, W.~G., \& {Baker}, J.~C. 1971, \apj, 170, 241

\bibitem[{{McClure} \& {van den Bergh}(1968)}]{McClure1968}
{McClure}, R.~D., \& {van den Bergh}, S. 1968, \aj, 73, 1008

\bibitem[{{Mentz} {et~al.}(2016){Mentz}, {La Barbera}, {Peletier},
  {Falc{\'o}n-Barroso}, {Lisker}, {van de Ven}, {Loubser}, {Hilker},
  {S{\'a}nchez-Janssen}, {Napolitano}, {Cantiello}, {Capaccioli}, {Norris},
  {Paolillo}, {Smith}, {Beasley}, {Lyubenova}, {Munoz}, \& {Puzia}}]{Mentz2016}
{Mentz}, J.~J., {La Barbera}, F., {Peletier}, R.~F., {et~al.} 2016, \mnras,
  463, 2819

\bibitem[{{Moran} {et~al.}(2005){Moran}, {Ellis}, {Treu}, {Smail}, {Dressler},
  {Coil}, \& {Smith}}]{Moran2005}
{Moran}, S.~M., {Ellis}, R.~S., {Treu}, T., {et~al.} 2005, \apj, 634, 977

\bibitem[{{Moran} {et~al.}(2007){Moran}, {Ellis}, {Treu}, {Smith}, {Rich}, \&
  {Smail}}]{Moran2007}
---. 2007, \apj, 671, 1503

\bibitem[{{Muratov} {et~al.}(2015){Muratov}, {Kere{\v{s}}}, {Faucher-
  Gigu{\`e}re}, {Hopkins}, {Quataert}, \& {Murray}}]{Muratov2015}
{Muratov}, A.~L., {Kere{\v{s}}}, D., {Faucher- Gigu{\`e}re}, C.-A., {et~al.}
  2015, \mnras, 454, 2691

\bibitem[{{Naiman} {et~al.}(2018){Naiman}, {Pillepich}, {Springel},
  {Ramirez-Ruiz}, {Torrey}, {Vogelsberger}, {Pakmor}, {Nelson}, {Marinacci},
  {Hernquist}, {Weinberger}, \& {Genel}}]{Naiman2018}
{Naiman}, J.~P., {Pillepich}, A., {Springel}, V., {et~al.} 2018, \mnras, 477,
  1206

\bibitem[{{Newman} {et~al.}(2013){Newman}, {Cooper}, {Davis}, {Faber}, {Coil},
  {Guhathakurta}, {Koo}, {Phillips}, {Conroy}, {Dutton}, {Finkbeiner}, {Gerke},
  {Rosario}, {Weiner}, {Willmer}, {Yan}, {Harker}, {Kassin}, {Konidaris},
  {Lai}, {Madgwick}, {Noeske}, {Wirth}, {Connolly}, {Kaiser}, {Kirby},
  {Lemaux}, {Lin}, {Lotz}, {Luppino}, {Marinoni}, {Matthews}, {Metevier}, \&
  {Schiavon}}]{Newman2013}
{Newman}, J.~A., {Cooper}, M.~C., {Davis}, M., {et~al.} 2013, The Astrophysical
  Journal Supplement Series, 208, 5

\bibitem[{{Niemiec} {et~al.}(2018){Niemiec}, {Jullo}, {Giocoli}, {Limousin}, \&
  {Jauzac}}]{Niemiec2018}
{Niemiec}, A., {Jullo}, E., {Giocoli}, C., {Limousin}, M., \& {Jauzac}, M.
  2018, arXiv e-prints, arXiv:1811.04996

\bibitem[{{Nomoto} {et~al.}(2006){Nomoto}, {Tominaga}, {Umeda}, {Kobayashi}, \&
  {Maeda}}]{Nomoto2006}
{Nomoto}, K., {Tominaga}, N., {Umeda}, H., {Kobayashi}, C., \& {Maeda}, K.
  2006, \nphysa, 777, 424

\bibitem[{{Nyland} {et~al.}(2013){Nyland}, {Alatalo}, {Wrobel}, {Young},
  {Morganti}, {Davis}, {de Zeeuw}, {Deustua}, \& {Bureau}}]{Nyland2013}
{Nyland}, K., {Alatalo}, K., {Wrobel}, J.~M., {et~al.} 2013, \apj, 779, 173

\bibitem[{{Pagel}(1997)}]{Pagel1997}
{Pagel}, B. E.~J. 1997, {Nucleosynthesis and Chemical Evolution of Galaxies}

\bibitem[{{Pettini} \& {Pagel}(2004)}]{PettiniPagel2004}
{Pettini}, M., \& {Pagel}, B. E.~J. 2004, \mnras, 348, L59

\bibitem[{{Pillepich} {et~al.}(2018){Pillepich}, {Springel}, {Nelson}, {Genel},
  {Naiman}, {Pakmor}, {Hernquist}, {Torrey}, {Vogelsberger}, {Weinberger}, \&
  {Marinacci}}]{Pillepich2018}
{Pillepich}, A., {Springel}, V., {Nelson}, D., {et~al.} 2018, \mnras, 473, 4077

\bibitem[{{Roberts-Borsani} \& {Saintonge}(2019)}]{RobertsBorsani2019}
{Roberts-Borsani}, G.~W., \& {Saintonge}, A. 2019, \mnras, 482, 4111

\bibitem[{{Rupke} {et~al.}(2017){Rupke}, {G{\"u}ltekin}, \&
  {Veilleux}}]{Rupke2017}
{Rupke}, D. S.~N., {G{\"u}ltekin}, K., \& {Veilleux}, S. 2017, \apj, 850, 40

\bibitem[{{Salpeter}(1955)}]{Salpeter1955}
{Salpeter}, E.~E. 1955, \apj, 121, 161

\bibitem[{{S{\'a}nchez-Bl{\'a}zquez} {et~al.}(2006){S{\'a}nchez-Bl{\'a}zquez},
  {Peletier}, {Jim{\'e}nez- Vicente}, {Cardiel}, {Cenarro},
  {Falc{\'o}n-Barroso}, {Gorgas}, {Selam}, \& {Vazdekis}}]{Sanchez2006}
{S{\'a}nchez-Bl{\'a}zquez}, P., {Peletier}, R.~F., {Jim{\'e}nez- Vicente}, J.,
  {et~al.} 2006, \mnras, 371, 703

\bibitem[{{Saracco} {et~al.}(2019){Saracco}, {La Barbera}, {Gargiulo},
  {Mannucci}, {Marchesini}, {Nonino}, \& {Ciliegi}}]{Saracco2019}
{Saracco}, P., {La Barbera}, F., {Gargiulo}, A., {et~al.} 2019, \mnras, 484,
  2281

\bibitem[{{Sarzi} {et~al.}(2016){Sarzi}, {Kaviraj}, {Nedelchev}, {Tiffany},
  {Shabala}, {Deller}, \& {Middelberg}}]{Sarzi2016}
{Sarzi}, M., {Kaviraj}, S., {Nedelchev}, B., {et~al.} 2016, \mnras, 456, L25

\bibitem[{{Spitoni} {et~al.}(2010){Spitoni}, {Calura}, {Matteucci}, \&
  {Recchi}}]{Spitoni2010}
{Spitoni}, E., {Calura}, F., {Matteucci}, F., \& {Recchi}, S. 2010, \aap, 514,
  A73

\bibitem[{{Taylor} \& {Kobayashi}(2016)}]{TaylorKobayashi2016}
{Taylor}, P., \& {Kobayashi}, C. 2016, \mnras, 463, 2465

\bibitem[{{Thomas} {et~al.}(2005){Thomas}, {Maraston}, {Bender}, \& {Mendes de
  Oliveira}}]{Thomas2005}
{Thomas}, D., {Maraston}, C., {Bender}, R., \& {Mendes de Oliveira}, C. 2005,
  \apj, 621, 673

\bibitem[{{Tremonti} {et~al.}(2004){Tremonti}, {Heckman}, {Kauffmann},
  {Brinchmann}, {Charlot}, {White}, {Seibert}, {Peng}, {Schlegel}, {Uomoto},
  {Fukugita}, \& {Brinkmann}}]{Tremonti2004}
{Tremonti}, C.~A., {Heckman}, T.~M., {Kauffmann}, G., {et~al.} 2004, \apj, 613,
  898

\bibitem[{{van Dokkum} {et~al.}(2010){van Dokkum}, {Whitaker}, {Brammer},
  {Franx}, {Kriek}, {Labb{\'e}}, {Marchesini}, {Quadri}, {Bezanson},
  {Illingworth}, {Muzzin}, {Rudnick}, {Tal}, \& {Wake}}]{VanDokkum2010}
{van Dokkum}, P.~G., {Whitaker}, K.~E., {Brammer}, G., {et~al.} 2010, \apj,
  709, 1018

\bibitem[{{Vincenzo} {et~al.}(2016){Vincenzo}, {Matteucci}, {Belfiore}, \&
  {Maiolino}}]{Vincenzo2016}
{Vincenzo}, F., {Matteucci}, F., {Belfiore}, F., \& {Maiolino}, R. 2016,
  \mnras, 455, 4183

\bibitem[{{Werk} {et~al.}(2014){Werk}, {Prochaska}, {Tumlinson}, {Peeples},
  {Tripp}, {Fox}, {Lehner}, {Thom}, {O'Meara}, {Ford}, {Bordoloi}, {Katz},
  {Tejos}, {Oppenheimer}, {Dav{\'e}}, \& {Weinberg}}]{Werk2014}
{Werk}, J.~K., {Prochaska}, J.~X., {Tumlinson}, J., {et~al.} 2014, \apj, 792, 8

\bibitem[{{Wetzel} {et~al.}(2013){Wetzel}, {Tinker}, {Conroy}, \& {van den
  Bosch}}]{Wetzel2013}
{Wetzel}, A.~R., {Tinker}, J.~L., {Conroy}, C., \& {van den Bosch}, F.~C. 2013,
  \mnras, 432, 336

\bibitem[{{Wilkinson} {et~al.}(2017){Wilkinson}, {Maraston}, {Goddard},
  {Thomas}, \& {Parikh}}]{Wilkinson2017}
{Wilkinson}, D.~M., {Maraston}, C., {Goddard}, D., {Thomas}, D., \& {Parikh},
  T. 2017, \mnras, 472, 4297

\bibitem[{{Young} {et~al.}(2011){Young}, {Bureau}, {Davis}, {Combes},
  {McDermid}, {Alatalo}, {Blitz}, {Bois}, {Bournaud}, {Cappellari}, {Davies},
  {de Zeeuw}, {Emsellem}, {Khochfar}, {Krajnovi{\'c}}, {Kuntschner},
  {Lablanche}, {Morganti}, {Naab}, {Oosterloo}, {Sarzi}, {Scott}, {Serra}, \&
  {Weijmans}}]{Young2011}
{Young}, L.~M., {Bureau}, M., {Davis}, T.~A., {et~al.} 2011, \mnras, 414, 940

\bibitem[{{Zahid} {et~al.}(2013){Zahid}, {Geller}, {Kewley}, {Hwang},
  {Fabricant}, \& {Kurtz}}]{Zahid2013}
{Zahid}, H.~J., {Geller}, M.~J., {Kewley}, L.~J., {et~al.} 2013, \apj, 771, L19

\end{thebibliography}
\end{document}